\documentclass[notitlepage,letterpaper,aps,prd,onecolumn,superscriptaddress,nofootinbib,longbibliography]{revtex4-1}
\usepackage{amsmath}
\usepackage{amsfonts}
\usepackage{amssymb}
\usepackage[latin1,utf8]{inputenc}
\usepackage{xcolor,cancel}
\usepackage[normalem]{ulem}
\newcommand{\stkout}[1]{\ifmmode\text{\sout{\ensuremath{#1}}}\else\sout{#1}\fi}
\RequirePackage{flushend}
\usepackage{multirow}

\usepackage{natbib}

\usepackage[colorlinks=true, pdfstartview = FitH, bookmarksopen = true, bookmarksopenlevel=2]{hyperref}
\hypersetup{pdftitle=Cosmological solutions and  growth index of matter perturbations in  $f(Q)$ gravity,pdfauthor={Wompherdeiki Khyllep, Jibitesh Dutta, Andronikos Paliathanasis}}

\usepackage[all]{hypcap}

\usepackage{latexsym}
\usepackage{graphicx}
\usepackage{color}
\usepackage{subfigure}

\begin{document}

\title{Cosmological solutions and  growth index of matter perturbations in  $f(Q)$ gravity}
\author{Wompherdeiki Khyllep}
\email{sjwomkhyllep@gmail.com}
\affiliation{Department of Mathematics, North-Eastern Hill University,
	Shillong, Meghalaya 793022, India}
\affiliation{Department of Mathematics,
	St.\ Anthony's College, Shillong, Meghalaya 793001, India}
\author{Andronikos Paliathanasis}
\email{anpaliat@phys.uoa.gr}
\affiliation{Institute of Systems Science, Durban University of Technology,
	Durban 4000, South Africa}
	\affiliation{Instituto de Ciencias F\'{\i}sicas y Matem\'{a}ticas, Universidad Austral de Chile, Valdivia 5090000, Chile}
\author{Jibitesh Dutta}
\email{jibitesh@nehu.ac.in}
\affiliation{Mathematics Division, Department of Basic Sciences and Social
	Sciences, North-Eastern Hill University,  Shillong, Meghalaya 793022, India}
\affiliation{Inter University Centre for Astronomy and Astrophysics, Pune
	411007, India }

\begin{abstract}
The present work studies one of Einstein's alternative formulations based on the non-metricity scalar $Q$ generalized as $f(Q)$ theory.  More specifically, we consider the power-law form of $f(Q)$ gravity i.e. $f(Q)=Q+\alpha\, Q^n$. Here, we analyze the behavior of the cosmological model at the background and perturbation level. Using the dynamical system analysis, at the background level,  the effective evolution of the model is the same as that of the $\Lambda$CDM  for $|n|<1$. Interestingly, the geometric component of the theory solely determined the late-time acceleration of the Universe. We also examine the integrability of the model by employing the method of singularity analysis. In particular, we find the conditions under which field equations pass the Painlev\'{e} test and hence possess the Painlev\'{e} property. While the equations pass the Painlev\'{e} test in the presence of dust for any value of $n$, the test is valid after the addition of radiation fluid only for $n<1$. Finally, at the perturbation level,  the behavior of matter growth index signifies a deviation of the model from the $\Lambda$CDM even for $|n|<1$.
\end{abstract}

\maketitle

\section{Introduction}\label{sec:intro}

A manifestation of gravity through spacetime curvature is
one of the most fundamental assumptions that stem from the equivalence principle. Geometrically, besides the curvature, torsion and non-metricity are also the fundamental objects associated with the manifold's connection determining the gravity \cite{BeltranJimenez:2019tjy}. Depending on the choice of connection, one can classify the theories of gravity into three classes. The first one uses the curvature,  the free torsion, and metric compatible connection, e.g. General Relativity (GR). The second class uses the metric compatible, curvature-free connection with torsion, e.g. Teleparallel Equivalent of GR \cite{aldro}. The last one uses a curvature and torsion-free connection, which is not metric compatible, e.g. Symmetric Teleparallel Equivalent of GR \cite{Nester:1998mp}. These three equivalent formulations based on the three different connections are commonly known as {\it The Geometrical Trinity of Gravity} \cite{BeltranJimenez:2019tjy}. Even if these three theories are equivalent at the level of field equations, their modifications may not be equivalent at the fundamental level \cite{Altschul:2014lua}.

A generalization of the symmetric teleparallel gravity which has gained recent attention is the $f(Q)$ gravity theory \cite{BeltranJimenez:2017tkd,Jimenez:2019ovq}. In this theory, one considers a flat and vanishing torsion connection where gravity
is described by a non-metricity scalar $Q$ and hence represents one of the geometrical equivalent formulations of GR. Interestingly, one can simplify the corresponding connection in partial
derivatives, which vanish for some coordinate choice called the coincident gauge. One of the essential features of the $f(Q)$ theory is that, unlike GR, we can also separate gravity from the inertial effects.  It is also worth mentioning that while the field equations in $f(R)$ gravity are fourth-order \cite{Sotiriou:2008rp},  they are of second-order in $f(Q)$ gravity, and hence, $f(Q)$ gravity is free from pathologies. Thus, the construction of this theory forms a novel starting point for various modified gravity theories. It also presents a  simple formulation in which self-accelerating solutions arise naturally in both the early and late Universe.

Various work in the literature suggest that the $f(Q)$ theory is one of the promising alternative formulations of gravity to explain cosmological observations \cite{re1,re2,re3,re4,re5}. Harko et. al. constructed a class of $f(Q)$ theories where $Q$ is coupled non-minimally to the matter Lagrangian. As a cosmological application they  show that it can represent an alternative approach to dark energy (DE) \cite{Harko:2018gxr}.   Observational constraints on the background behavior of several $f(Q)$ models have been performed by testing against various current background data such as  Type Ia Supernovae, Pantheon data, Hubble data, etc.  \cite{Lazkoz:2019sjl,Ayuso:2020dcu}. These studies conclude that viable $f(Q)$ models correspond to model parameters' values which resemble the GR-based model viz. the Lambda Cold Dark Matter ($\Lambda$CDM) model.  Mandal et. al. analyzed the energy conditions to restrict the parameters of the power-law and logarithmic $f(Q)$ models compatible with the observed behavior of the Universe \cite{Mandal:2020lyq}. In Ref. \cite{Jarv:2018bgs}, J\"{a}rv et. al. introduced a  class of theories in which a scalar field and non-metricity scalar $Q$  coupled non-minimally. They found that such a class of theories is related to the $f(Q)$ theory. On perturbing around the background Friedmann-Lema\^{i}tre-Robertson-Walker
spacetime,  unlike $f(T)$ gravity models, strong coupling issues are absent in the case of $f(Q)$ gravity \cite{Jimenez:2019ovq,Golovnev:2018wbh}.

The interesting cosmological behavior of the $f(Q)$ theory at the background level motivates us to investigate its global dynamics from a dynamical system perspective. Dynamical system tools have been extensively used in the context of cosmology (see \cite{Khyllep:2021yyp,Dutta:2020uha,Alho:2020cdg,Leon:2020cge,Giacomini:2020zmv,Paliathanasis:2020wjl,Christodoulidis:2019jsx,Coley:2019tyx,Basilakos:2019dof,Cid:2017wtf} for a few related recent work and \cite{Bahamonde:2017ize} for review). However, one of the drawbacks of the dynamical system approach is that the resulting dynamics depend on the choice of variables. The absence	of  interesting cosmological solutions does not always imply the inability of the theory to describe such solutions. It may be because the associated dynamical system  cannot capture the desired dynamics with the specific choice of variables.  The importance of choice of variables has been also highlighted in the study of $f(R)$ gravity \cite{Carloni:2015jla,Alho:2016gzi} and $f(T)$ gravity \cite{Hohmann:2017jao}.  Another drawback is that the dynamical system analysis cannot provide sufficient information on the evolution far from the critical points. Therefore, if one obtains the analytical solutions, one might even determine the dynamics that the dynamical system analysis cannot explain.  Most cosmological equations are nonlinear; therefore, one usually prefer numerical tools to solve them. The knowledge about integrability of the dynamical system is important to relate numerical solutions and the real solutions of the system \cite{mtsa1}. Therefore, the determination of the analytical solutions for the field equations is crucial to study the integrability of the system. Such analysis will provide a preliminary investigation on the viability of a given theory.

The solution of a differential equation usually refers to an explicit function connecting the dependent and independent variables of the differential equation. However, this is not a unique way to express the solution of differential equations. Alternatively, one can write the differential equation into an algebraic equation with the use of similarity transformations. The use of such similarity transformations is true when there exist a sufficient number of invariant functions or first integrals. The latter definition of integrability is mainly related to the concept of symmetry. An alternative approach describing integrability based on movable singularities was established by Kovalevskaya \cite{sof}. This pioneering approach was applied to determine the third integrable case of Euler's equations for a spinning top. Based on this approach, the French School of
Painlev\'{e} at the beginning of the last century established the method of singularity analysis \cite{pain1,pain2,pain3,pain4}.  In this approach, the given differential equation is deemed integrable if it possesses the Painlev\'{e} property. The latter property is directly related to the existence of a Laurent expansion about a movable polelike singularity in the complex plane,  describing the relations between the dependent and independent variables of the differential equations. Hence the solution is expressed in terms of power series. In the last few years, the singularity analysis has been applied extensively to investigate the integrability of gravitational models with an emphasis on modern cosmology \cite{Cotsakis_1994,Christiansen_1995,Demaret_1996,Helmi:1997mj,Miritzis:2000js,Leach:2001aw,Leon:2018lnd,Basilakos:2018xjp,Paliathanasis:2016tch,Paliathanasis:2016vsw,Cotsakis:2013aqa,Paliathanasis:2017apr,Paliathanasis:2019qch}.  The existence of a movable singularity for the cosmological field equations can be related to the existence of cosmological singularities or with the dominant factor from the fluid components. Suppose the field equations possess the Painlev\'{e} property. In that case, the analytic solution can be written in terms of Laurent expansion or specifically with the use of Puiseux series, where the dominant term can be seen as an asymptotic solution. Thus, it is possible to extract information for the existence of movable singularities from the nature of the asymptotic solutions of the field equations. Such a discussion can be found in \cite{asze} where the asymptotic solutions  of the Szekeres system are related with the dominant terms  of the Painlev\'{e} Series describing the analytic solution for the gravitational model. Furthermore, we can determine the stability properties of the solution from  the nature of the series. In the present work, we shall attempt to determine the cosmological analytical solutions of the power-law model of $f(Q)$ theory motivated by previous work on singularity analysis in cosmology.

After examining the gravity theory at the background level, the next logical step is to test its viability at the perturbation level.  The study on the growth rate of matter perturbations is an effective approach to estimate the distribution of matter in the Universe \cite{Dutta:2017wfd} and also to theoretically differentiate various gravity theories \cite{Ishak:2018his,bb1,Khyllep:2019odd}. For instance, the growth index is approximately constant throughout the evolution of GR-based DE models. However, there is a significant variation in the value of the growth index in the case of modified gravity theories. It is important to note that the growth index of matter perturbations is one of the observational tools to study the matter's growth history of a given model \cite{peeblesbook}. Therefore, studying the evolution of  matter perturbations will allow us to draw an overall impact of the $f(Q)$ gravity at the cosmological level.

The background behavior of the specific $f(Q)$ model (i.e. $f(Q)=Q+\alpha\,Q^{\frac{1}{2}}$)  mimics that of the $\Lambda$CDM. However, it shows deviation at the perturbative level by testing against redshift space distortion data \cite{Barros:2020bgg}. Further, a deviation of $f(Q)$ model from the $\Lambda$CDM  at the linear perturbation level is observed by analyzing their prediction towards the matter power spectrum, lensing power spectrum, and an enhanced integrated-Sachs-Wolfe effect. Thus it is imperative to investigate the evolution behavior of the $f(Q)$ model at the perturbation level by analyzing the nature of the growth index of matter perturbations. Therefore, in this work we shall investigate the dynamics of the power-law $f(Q)$ model at the background level and the linear growth index of matter perturbations.

The work plan is as follows: In Sec. \ref{sec:f_Q_comology}, we present the basic cosmological equations of the general $f(Q)$ theory. We performed the dynamical system analysis of the power-law model of $f(Q)$ theory in Sec. \ref{sec:DSA}. We follow this by the determination of the analytic solutions of the model using the singularity analysis method in Sec. \ref{sec:SAM} for two cases: (a) dust fluid only in subsection \ref{ssec:dust}, (b) dust fluid along with radiation in subsection \ref{ssec:dust_rad}. We then investigate the implications of the growth of linear matter perturbations within the sub-horizon scale in Sec. \ref{sec:GI}.  Finally, we draw  our conclusion in Sec. \ref{sec:conc}.

\section{$f(Q)$ cosmology}\label{sec:f_Q_comology}
 In the present work, we shall consider a modified
 gravity theory in which the fundamental object is the
 non-metricity tensor given by \cite{BeltranJimenez:2017tkd}
 \begin{equation}
 Q_{\alpha \mu \nu }=\nabla _{\alpha }g_{\mu \nu }\,,
 \end{equation}
 where $g_{\mu\nu}$ is the metric. The two independent traces of $Q_{\alpha\mu\nu}$ are
 \begin{equation}
 Q_{\alpha}=Q_{\alpha }{}^{\mu }{}_{\mu }\,,\quad \tilde{Q}_{\alpha }=Q^{\mu
 }{}_{\alpha \mu }\,.
 \end{equation}
 The invariant non-metricity scalar is defined as a contraction of $Q_{\alpha
 	\mu \nu }$ given by
 \begin{equation}
 Q=-Q_{\alpha \mu \nu }P^{\alpha \mu \nu}\,,
 \end{equation}
 where $P^{\alpha \mu \nu}$ is the non-metricity conjugate and
 \begin{eqnarray}
 4P^{\alpha }{}_{\mu \nu } &=& -Q^{\alpha }{}_{\mu \nu } + 2Q_{(\mu \phantom{\alpha}\nu )}^{\phantom{\mu}\alpha } - Q^{\alpha }g_{\mu \nu } - \tilde{Q}^{\alpha }g_{\mu \nu }-\delta _{(\mu }^{\alpha }Q_{\nu )}\,.
 \label{super}
 \end{eqnarray}
 Using the non-metricity scalar, the action of the $f(Q)$ gravity is
 given by \cite{BeltranJimenez:2017tkd}
 \begin{equation}  \label{qqm}
 S=\int \left[\frac{1}{2}f(Q)+\mathcal{L}_m\right]\sqrt{-g}~d^4x,
 \end{equation}
 where $f(Q)$ is an arbitrary function of the scalar $Q$, $g$ is the
 determinant of  $g_{\mu\nu}$ and $\mathcal{L}_m$ is the matter
 Lagrangian density.
 
 \noindent On varying the action (\ref{qqm}) with respect to the metric, one obtains
 the corresponding Einstein's field equations
 \begin{equation*}  \label{EFE}
 \frac{2}{\sqrt{-g}}\nabla _{\alpha }\left( \sqrt{-g}f_{Q}P^{\alpha }{}_{\mu
 	\nu }\right) +\frac{1}{2}g_{\mu \nu }f+f_{Q}\left( P_{\mu \alpha \beta
 }Q_{\nu }{}^{\alpha \beta }-2Q_{\alpha \beta \mu }P^{\alpha \beta }{}_{\nu
 }\right) =-T_{\mu \nu }\,,
 \end{equation*}%
 where $f_{Q}=\frac{df}{dQ}$ and $T_{\mu \nu }=-\frac{2}{\sqrt{-g}}\frac{\delta \left( \sqrt{-g}\mathcal{L}_{m}\right) }{\delta g^{\mu \nu }}$. We
 assume that the matter is a perfect fluid whose energy-momentum tensor $T_{\mu \nu }$ is given by
 \begin{equation*}
 T_{\mu \nu }=(\rho +p)u_{\mu }u_{\nu }+pg_{\mu \nu }\,,
 \end{equation*}%
 where $u_{\mu }$ is the four-velocity satisfying the normalization condition
 $u_{\mu }u^{\mu }=-1$, $\rho $ and $p$ are the energy density and pressure
 of a perfect fluid respectively. Under the homogeneous and isotropic
 universe described by the Friedmann-Lema\^{i}tre-Robertson-Walker metric
 \begin{equation}
 ds^{2}=-dt^{2}+a^{2}(t)\delta _{\mu \nu }dx^{\mu }dx^{\nu }\,,~~~~~~~(\mu
 ,\nu =0,1,2,3)
 \end{equation}%
 the non-metricity scalar is given by $Q=6H^{2}$, where  $H=\frac{\dot{a}}{a}$ is the Hubble parameter with $a(t)$ denoting  scale factor and the upper dot denotes derivative with respect to the coordinate time  $t$. On taking  $f(Q)=Q+F(Q)$, the corresponding field equations can be written as
 \begin{eqnarray}
 3H^{2}&=&\rho+\frac{F}{2}-QF_Q \,,  \label{FRDEQ11} \\
 \left(2Q F_{QQ}+F_Q+1\right) \dot{H}+\frac{1}{4} \left(Q+2QF_Q-F\right)&=&-2p\,.  \label{FRDEQ22}
 \end{eqnarray}%
 Here, we consider the case that the Universe is filled with
 dust  and radiation fluids, therefore
 \begin{equation}
 \rho =\rho _{m}+\rho _{r},~~~~p=\frac{1}{3}\rho _{r}\,,
 \end{equation}%
 where $\rho _{m}$ and $\rho _{r}$ are the energy densities of dust and radiation, respectively. Then from equations \eqref{FRDEQ11} and \eqref{FRDEQ22}, we get
 \begin{eqnarray}
 H^{2} &=&\frac{1}{3}\left( \rho _{m}+\rho _{r}+\rho _{\mathrm{de}}\right) \,,
 \label{FRDEQ1} \\
2\dot{H}+3H^2&=& -\frac{\rho_r}{3}-p_{\rm de}\,,  \label{FRDEQ2}
 \end{eqnarray}%
 where $\rho _{\mathrm{de}}$ and $p_{\rm de}$ are respectively the DE's density and pressure contribution due to the geometry given by
 \begin{eqnarray}
 \rho _{\rm{de}}&=&\frac{F}{2}-QF_{Q}\,,\\
 p_{\rm de}&=&2 \dot{H} (2QF_{QQ}+F_Q)-\rho_{\mathrm{de}}\,.
 \end{eqnarray}
 Therefore, the equation of state  due to DE is given by
 \begin{eqnarray}
 w_{\rm de}&=&\frac{p_{\rm de}}{\rho_{\rm de}}= -1+\frac{4\dot{H}(2QF_{QQ}+F_Q)}{F-2QF_Q}\,.
 \end{eqnarray}
 Additionally, assuming that matter and
 radiation are not interacting each other, the conservation equation of the
 energy-momentum tensor for pressureless matter and radiation can be
 respectively written as
 \begin{equation}
 \dot{\rho}_{m}+3H\rho _{m}=0\,,~~~~\dot{\rho}_{r}+4H\rho _{r}=0\,.
 \label{matcons}
 \end{equation}%
 From the above equations, we  obtain $\rho _{m}=\rho _{m0}a^{-3}$
 and $\rho _{r}=\rho _{r0}a^{-4}$ where $\rho_{m0}$, $\rho_{r0}$  denote
 the matter and radiation energy density at the present time.  To  better understand the evolution of energy densities, we respectively  introduce the energy density parameters of a pressureless matter, radiation, and DE as
 \begin{equation}
 \Omega _{m}=\frac{\rho _{m}}{3H^{2}},~~~\Omega _{r}=\frac{\rho _{r}}{3H^{2}} ,~~~\Omega _{\mathrm{de}}=\frac{\rho _{\mathrm{de}}}{3H^{2}}\,.
 \end{equation}%
 The Friedman equation  \eqref{FRDEQ1} relates the above-defined quantities  as
 \begin{equation}
 \Omega _{m}+\Omega _{r}+\Omega _{\mathrm{de}}=1\,.  \label{constraint}
 \end{equation}%
 From \eqref{constraint}, one can define the matter dominated universe as a
 scenario where $\Omega _{m}=1,\Omega _{r}=0,\Omega _{\mathrm{de}}=0$.
 Similarly, one can also define radiation dominated universe or DE
 dominated universe, when $\Omega _{r}$ or $\Omega _{\mathrm{de}}$ dominates
 over the others, respectively. From equations \eqref{FRDEQ1} and \eqref{FRDEQ2}, we can also define the effective energy density $\rho _{\mathrm{eff}}$ and
 effective pressure $p_{\mathrm{eff}}$ respectively as
 \begin{eqnarray}
 \rho _{\mathrm{eff}} &=&\rho _{m}+\rho _{r}+\frac{F}{2}-QF_{Q}\,, \\
 p_{\mathrm{eff}} &=&\frac{\rho _{r}}{3}+\left( QF_{Q}+\frac{F}{2}\right)
 -\left( \rho _{m}+\frac{4}{3}\rho _{r}\right) \left( \frac{2QF_{QQ}+F_{Q}}{ 	2QF_{QQ}+1+F_{Q}}\right) \,.
 \end{eqnarray}%
 Therefore, the effective equation of state $w_{\mathrm{eff}}$ is given by
 \begin{equation}
 w_{\mathrm{eff}}=\frac{p_{\mathrm{eff}}}{\rho _{\mathrm{eff}}}=-1+\frac{\Omega _{m}+\frac{4}{3}\Omega _{r}}{2QF_{QQ}+1+F_{Q}}\,.
 \end{equation}%
 Similarly, we can define the deceleration  parameter $q$ which is directly related to $w_{\rm eff}$ as
 \begin{eqnarray}
 q&=& -1-\frac{\dot{H}}{H^2}=\frac{1+3 w_{\rm eff}}{2}\,.
 \end{eqnarray}
The deceleration parameter is of fundamental importance as it describes whether the universe undergoes acceleration ($q<0$ or $w_{\mathrm{eff}}<-\frac{1}{3}$) or deceleration  ($q>0$ or $w_{\mathrm{eff}}>-\frac{1}{3}$). To obtain a qualitative information on the solution's dynamical features of the
 system of cosmological equations, in the next section, we shall
 analyze the dynamics using the dynamical system techniques.
 
 \section{Dynamical system analysis}\label{sec:DSA}
 To analyze the dynamics of the $f(Q)$ model, we shall
 transform the equations of motion \eqref{FRDEQ11}, \eqref{FRDEQ22} and  \eqref{matcons} into an autonomous system of the first-order differential
 equations using the following dimensionless variables:
 \begin{eqnarray}  \label{var}
 x=\frac{2Q F_Q-F}{6H^2},\quad \quad y=\frac{\rho_r}{3H^2}\,.
 \end{eqnarray}
 Basically, by referring from the cosmological equations \eqref{FRDEQ11}, \eqref{FRDEQ22} and \eqref{matcons}, we have four dynamical variables viz.~$H, \rho_m, \rho_r$
 and $\rho_{\rm de}$. As we have considered the usual $H$-normalized variables, the variable $H$ is combined with other variables, so we are left with three
 variables. However, the remaining variables are connected by relation 
 \eqref{constraint}, therefore, we are left with only two independent
 variables which are expressed as $x$ and $y$. As favored by observations, we consider an expanding universe i.e. $H>0$ and hence, the above variables are well-defined.
 
 Using the above variables, the cosmological equations \eqref{FRDEQ11}, \eqref{FRDEQ22} and \eqref{matcons} can be transformed
 into the following dynamical system:
 \begin{eqnarray}
 x^{\prime } &=&2\frac{\dot{H}}{H^{2}}\Big[\left( F_{Q}-2QF_{QQ}\right) -x\Big]\,,
 \label{eq:xdys} \\
 y^{\prime } &=&-2y\left( 2+\frac{\dot{H}}{H^{2}}\right) \,,  \label{eq:ydys}
 \end{eqnarray}%
 where  prime denotes a derivative with respect to $\ln a$ and
 \begin{equation*}
 \frac{\dot{H}}{H^{2}}=-\frac{1}{2}\frac{3-3x+y}{2QF_{QQ}+F_{Q}+1}\,.
 \end{equation*}%
 To close the above system, one has to specify the function $F(Q)$. In case
 the system cannot be closed, one has to introduce additional variables which
 increase the dimension of the system. In this work, we will focus on a
 power-law form of function $F$ given by
 \begin{eqnarray}
 F(Q)=\alpha \,Q^n\,,
 \end{eqnarray}
where $\alpha $ and $n$ are dimensionless parameters.   We remark here that for $n=0$, the model reduces to the standard $\Lambda$CDM model with the quantity $\frac{\alpha}{2}$ playing the role of the cosmological constant \cite{Lazkoz:2019sjl,Barros:2020bgg}. The case $n=1$ is
 equivalent to the Symmetric Teleparallel Equivalent of General Relativity
  subject to the rescaling of Newton's gravitational constant by a
 factor of $\alpha +1$ \cite{Jimenez:2019ovq}. However, modification from the
 GR evolution occurs at low curvatures regime for $n<1$ and modification at
 high curvatures regime occurs for $n>1$. Hence, while models with $n>1$ will
 be applicable for the early Universe, models with $n<1$ will be applicable
 to the late-time DE dominated Universe. Therefore, we shall focus
 on the case where $n\neq 1$.  For this example, we have
 \begin{equation}
 F_Q+2 Q F_{QQ}=nx\,,
 \end{equation}
 and hence, the system \eqref{eq:xdys}-\eqref{eq:ydys} can be rewritten as
 \begin{eqnarray}
 x'&=&\frac{(1-n)  (3-3x+y)x}{nx+1}\,, \label{eq:xdys_ex} \\
 y'&=&- \frac{y \left[(4n+3) x-y+1\right]}{nx+1}\,.\label{eq:ydys_ex}
 \end{eqnarray}
 Further,	 we can rewrite  $\Omega_r$, $\Omega_{\rm de}$,  $\Omega_m$, $w_{\rm eff}$ and $w_{\rm de}$ in terms of variables $x, y$ as
 \begin{gather*}
 \Omega _{r}=y\,,~~~~~~~~~~~~~~~~\Omega _{\mathrm{de}}=x\,,~~~~~~~~~~~~~~\Omega _{m}=1-x-y\,,~~~~~~~~~~~~~~~~~\\
 w_{\mathrm{eff}}=-\frac{1}{3}\frac{3\,nx+3\,x-y}{nx+1}\,,~~~~~w_{\rm de}=-1-\frac{2n}{3}\frac{\dot{H}}{H^2}=-1+\frac{n(3-3x+y)}{3(nx+1)}\,.
 \end{gather*}%
 Under a physical condition $0\leq \Omega _{m},\Omega _{r}\leq 1$, the phase
 space of the system \eqref{eq:xdys_ex}-\eqref{eq:ydys_ex} is given by
 \begin{equation*}
 \Psi =\left\{ (x,y)\in \mathbb{R}^{2}\big|~0\leq x+y\leq 1,0\leq y\leq 1, -y\leq x\leq 1-y \right\} \,.
 \end{equation*}
 
 To analyze the cosmological dynamics of the model, we extract the critical
 points of the system \eqref{eq:xdys_ex}-\eqref{eq:ydys_ex} by solving the equations $
 x^{\prime }=0$ and $y^{\prime }=0$.  The system contains three critical points presented in Table \ref{tab:c_pts_general} whose stability property depends on the value of $n$. In what follows, we describe the stability nature of each critical point by examining the eigenvalues corresponding to the Jacobian matrix of a system \eqref{eq:xdys_ex}-\eqref{eq:ydys_ex} at each point.
 
 \begin{table*}[!ht]
		\centering
		\small
		\caption{Critical points of the system \eqref{eq:xdys_ex}-\eqref{eq:ydys_ex}.}
		\label{tab:c_pts_general}
			\begin{tabular}{cccccccc}
				\hline\hline
				Point$(x,y)$~~~ &Existence&~~~ $\Omega_m$~~~ &~~     $\Omega_{r}$ ~~& ~~  $\Omega_{\rm de}$ ~~& $w_{\rm eff}$ ~~&~~Acceleration~~~ & Stability   \\
				\hline
				\multirow{3}*{$A(0,1)$}~~ & ~~\multirow{3}*{Always}&    \multirow{3}*{$0$}     &      \multirow{3}*{$ 1$}   & \multirow{3}*{$0$}&    \multirow{3}*{$\frac{1}{3}$}  & \multirow{3}*{No} & unstable node if $n<1$\\
				&&&&&&& saddle if $n>1$\\ &&&&&&&
				\\%[1.5ex]
				\multirow{3}*{$B(0,0)$} ~~&~~ \multirow{3}*{Always} &     \multirow{3}*{$1$}     &      \multirow{3}*{$0$}   & \multirow{3}*{$0$}&    \multirow{3}*{$0$}  & \multirow{3}*{No} & saddle if $n<1$\\
				&&&&&&& stable node if $n>1$\\ &&&&&&&
				\\%[1.5ex]
				\multirow{3}*{$C(1,0)$}~~ &~~ \multirow{3}*{$n \neq -1$} &     \multirow{3}*{$0$}     &      \multirow{3}*{$0$}   & \multirow{3}*{$1$} &    \multirow{3}*{$-1$} & \multirow{3}*{Always} & stable node if $n<1$\\
				&&&&&&& saddle if $n>1$\\ &&&&&&&
				\\%[1.5ex]
								\hline\hline
		\end{tabular} 
				\end{table*}

 \begin{itemize}
 	\item \textit{Point A}$(0,1)$ corresponds to a decelerated, radiation dominated universe (
 	$\Omega_r=1, w_{\mathrm{eff}}=\frac{1}{3}$). In this case, the eigenvalues are $1$, $4(1-n)$. Therefore,
 	this point is an unstable node when $n<1$, saddle if $n>1$.
 	
 	\item \textit{Point B}$(0,0)$ corresponds to a decelerated,  matter dominated universe ($
 	\Omega_m=1, w_{\mathrm{eff}}=0$). As the eigenvalues of this point are $-1$, $3(1-n)$, therefore, this point is
 	a stable node when $n>1$, saddle if $n<1$.
 	
 	\item \textit{Point C}$(1,0)$ corresponds to an accelerated, DE dominated
 	universe ($\Omega_{\mathrm{de}}=1, w_{\mathrm{eff}}=-1$). Note that
 	denominator of the right hand side of the equations \eqref{eq:xdys_ex}-
 	\eqref{eq:ydys_ex} is $nx+1$, therefore, this point does not exist for $n=-1$.
 	The eigenvalues evaluated at this point 	are $-4$, $\frac{3(n-1)}{n+1}$. Therefore, this point is a stable node when $n<1$, saddle if $n>1$.
 \end{itemize}
 
 \begin{figure}[tbp]
 	\centering
 	\subfigure[]{
 		\includegraphics[width=7cm, height=7.2cm]{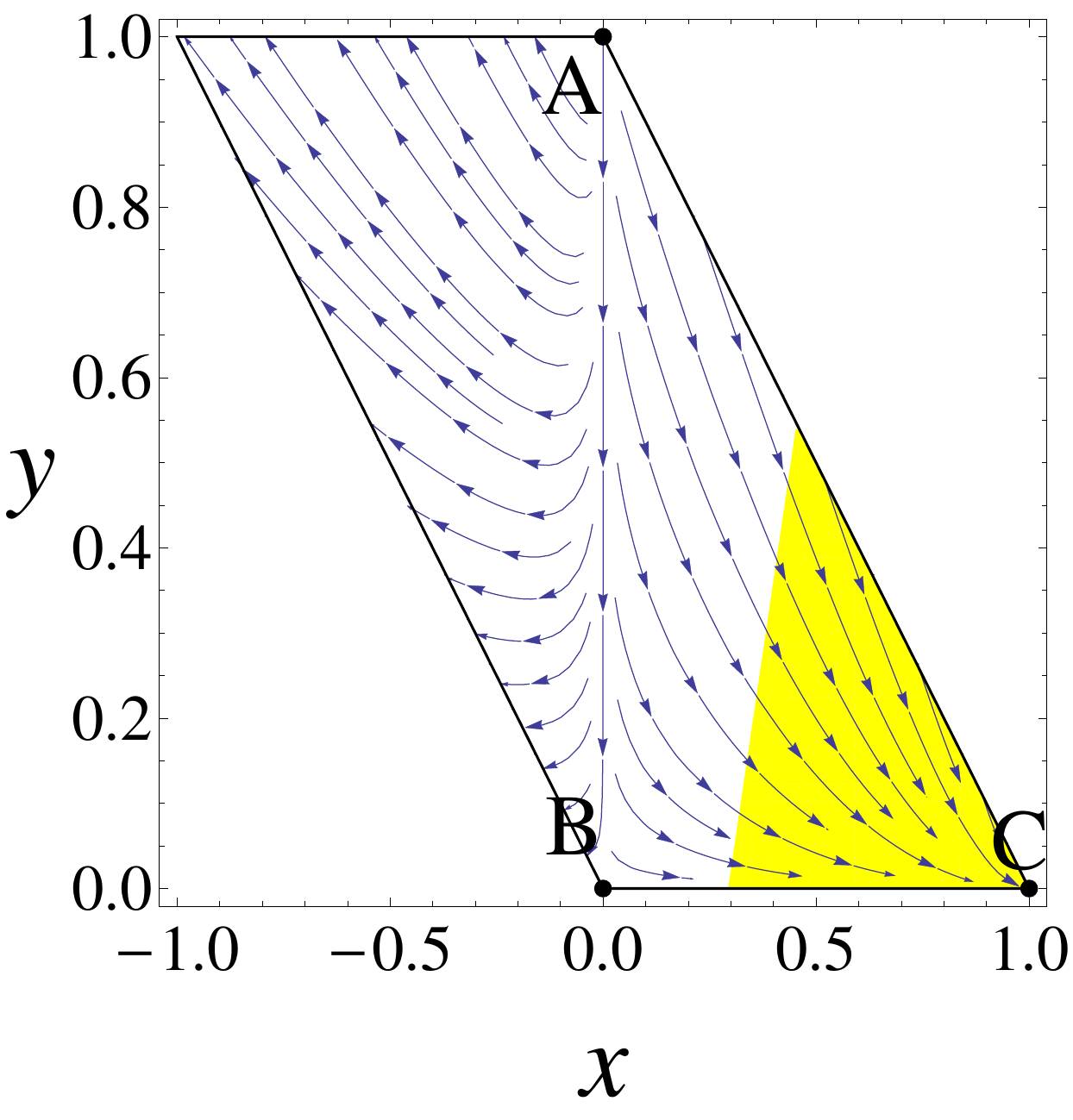} \label{fig:phase_plot}}
 	\qquad
 	\subfigure[]{
 		\includegraphics[width=7cm, height=7cm]{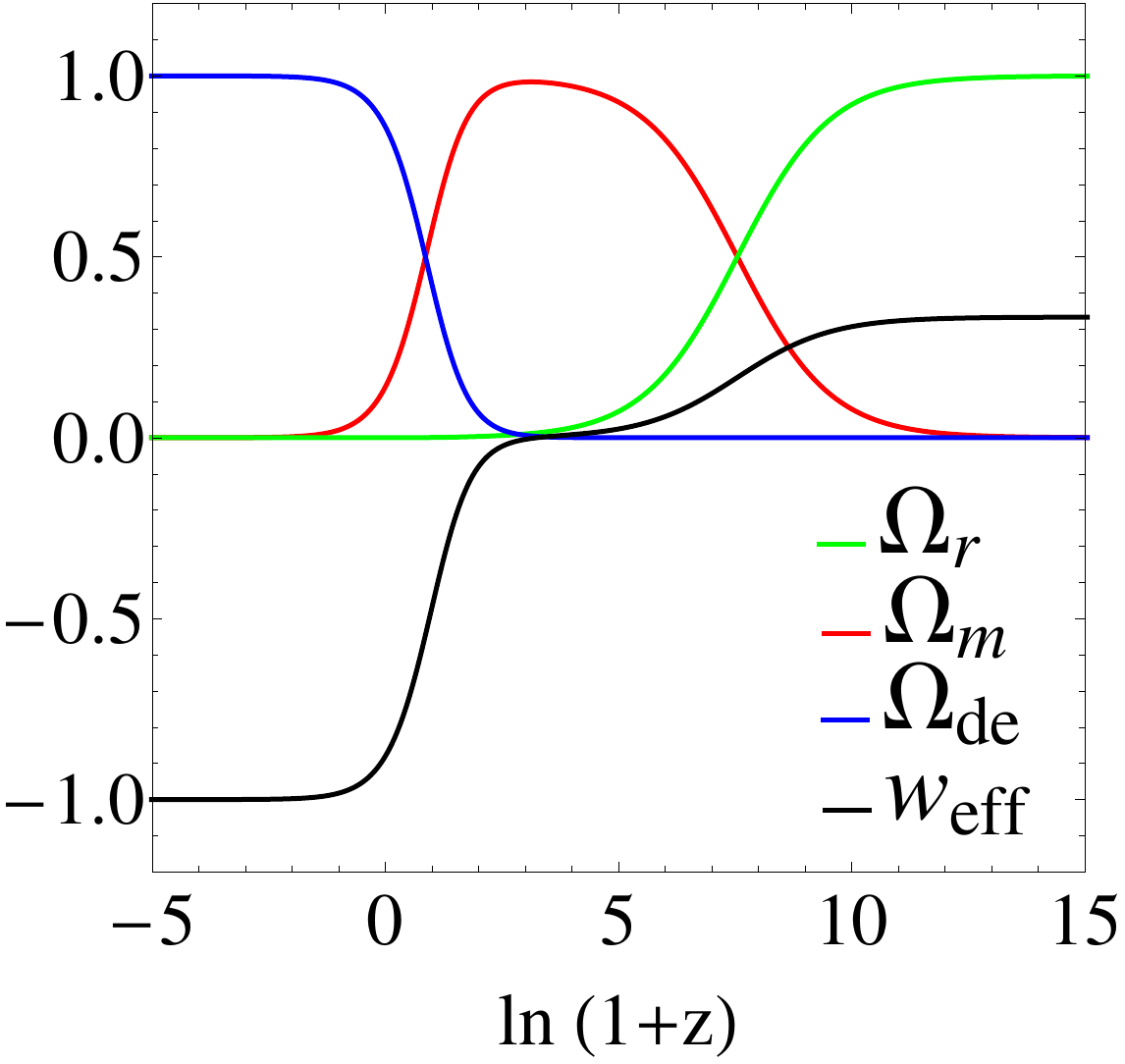} \label{fig:bcg_para} }
 	\caption{(a) Phase portrait of the system \eqref{eq:xdys_ex}-\eqref{eq:ydys_ex}
 		 with $n=0.2$. The shaded
 		area represents the region of accelerated expansion. (b) Time evolution of
 		$\Omega_{r}$, $\Omega_{m}$, $\Omega_{\rm de}$ and $w_{\rm eff}$. Here, $z$ denotes redshift given by $z=\frac{a_0}{a}-1$ where	$a_0 = 1$ is the scale factor at the present time.}
 	\label{fig:bcg_plot}
 \end{figure}
 
 Indeed, for $n=0$ one recover the $\Lambda$CDM model. From the stability behavior of critical points, we see that for $-1<n<1$, the Universe evolves from a radiation dominated epoch towards a matter dominated epoch and eventually settles to an accelerated DE dominated epoch. The phase space evolution describing the transition of the Universe along with the evolution of the cosmological parameters is given in Fig. \ref{fig:bcg_plot}. Additionally, we also require a fine-tuning of the initial conditions to have a long matter dominated epoch. Otherwise, we obtain an unusual early behavior where the Universe evolves directly from radiation dominated epoch to DE dominated one. So, for $|n|<1$, the present model's background dynamics are consistent with the present observational data, with the effective evolution resembles that of the $\Lambda$CDM. It is important to note that late-time acceleration is solely due to the geometric sector of the action.
 
\section{Analytical solutions by Singularity analysis method}\label{sec:SAM}
To determine the analytical solution of the cosmological equations, we apply
the singularity analysis technique. The method of singularity analysis has
been extensively applied to various cosmological models based on the
standard GR \cite{Cotsakis_1994,Christiansen_1995,Demaret_1996,Helmi:1997mj,Miritzis:2000js,Leach:2001aw,Leon:2018lnd,Basilakos:2018xjp}
or modified gravity theories \cite{Paliathanasis:2016tch,Paliathanasis:2016vsw,Cotsakis:2013aqa,Paliathanasis:2017apr,Paliathanasis:2019qch}. In the present work, we focus on a systematic technique of singularity analysis known as the Ablowitz-Ramani-Segur (ARS) algorithm \cite{ars1,ars2,ars3}. The algorithm allows us to determine whether a given
differential equation passes the Painlev\'{e} test and hence possesses the Painlev\'e property i.e., the solution can be
written as a Painlev\'{e} series (or Laurent expansion) around a movable
singularity.

Consider the differential equation $H\left( t,a,\dot{a},\ddot{a} \left( t\right) ,...\right) =0,$ where $a\left( t\right) $ is the dependent variable and $t$ is the independent variable. The application of the ARS algorithm is summarized in the following three steps:

The first step of the ARS algorithm is to determine whether a
movable singularity exists. We substitute $a\left( t\right) =a_{0}\tau ^{p}$
in the differential equation in which $\tau =t-t_{0}$	and $t_{0}~$is the location of the putative singularity. The dominant powers which share a common scale symmetry are selected, and we require them to be the dominant behaviour of the original differential equation. Hence, the coefficient $a_{0}$ and the exponent $p$ are determined. The exponent $p$ should be a negative value for the singularity to be a pole. However,
nowadays, exponent $p$ can be a fractional number, even positive
ones, as the derivative of a positive fractional exponent eventually give a negative exponent and so a singularity.

The second step of the ARS algorithm is determining the resonances, which provides the integration constants' location for the differential equation. In order to do that, we substitute $a\left( t\right)
=a_0 \tau ^{p}+m\tau ^{p+r}$ into the dominant terms of the equation and collect the terms linear in $m$ as that is where the coefficient firstly enters the expansion. If the multiplier of $m$ is zero, the value of $m$ is arbitrary. The coefficient is a polynomial in $r$. We then equate the coefficient to zero to obtain the values of $r$. One of the solutions must be $-1$, which is associated with the moveable singularity at $t_{0}$.

Finally, the third step in the ARS algorithm is to perform the consistency test by substituting  the Painlev\'{e} series to the original equation.  We emphasize that the step of the Painlev\'e series is determined by the leading order term and the resonances.  The idea is to check whether the series is indeed a true solution.

Note that the consistency test determines the coefficients of the Painlev\'{e} series. The nature of resonances determines the type of series. The series can be expressed as a right Laurent expansion; for negative resonances, the series is left Laurent expansion; otherwise, the series is a mixed Laurent expansion. In what follows, we perform the analysis for the power-law model in the presence of (a) dust fluid only, (b) dust with radiation fluid.

\subsection{Dust fluid}\label{ssec:dust}

For performing the first step of singularity analysis, we
substitute $a(\tau )=a_{0}\tau ^{\sigma }$ in \eqref{FRDEQ22} (taking $\rho
_{r}=0$ and hence right hand side of \eqref{FRDEQ22} vanishes), we then obtain

\begin{equation}
-2\,{\tau }^{-2\,n}\alpha \,{\sigma }^{2\,n-1}{6}^{n-1}\left( 2\,n-1\right)
\left( 2\,n-3\,\sigma \right) +2\,\tau ^{-2}\,\sigma \left( 3\,\sigma
-2\right) =0\,,  \label{FRDEQSA}
\end{equation}%
where $\tau =t-t_{0}$, $t_{0}$ is a constant of integration which determine
the position of singularity. Then, we search for the leading terms which
determine the value of $\sigma $. As discussed earlier, we focus our study
to two different cases $n<1$ and $n>1$.

\subsubsection{$n<1$ case}

For the case $n<1$, it follows from the equation \eqref{FRDEQSA}, that the leading order behavior is obtained from the term $2\,\tau^{-2}\,\sigma
\left( 3\, \sigma-2 \right)$. Therefore, if the leading-order behavior
describes the solution at the singularity then we have $\left( 3\, \sigma-2 \right)=0$ i.e. $\sigma=\frac{2}{3}$, which is independent of $n$ and $a_0$ is arbitrary. The leading term $a(\tau)=a_0 \tau^{\frac{2}{3}}$ implies that near singularity i.e. $t \to t_0$, we have $a(t) \to 0$ but the derivatives of $a(t)$ diverge.

Next in order to find the resonances $r$, we substitute $a(\tau )=a_{0}\tau^{\frac{2}{3}}+m\tau ^{\frac{2}{3}+r}$ in \eqref{FRDEQ22} and linearize around $m=0$. From the remaining terms, we solve $r$ from the coefficient of the leading order term and obtain an equation
\begin{equation}
r(r+1)=0\,,
\end{equation}%
giving two solutions $r_{1}=-1$ and $r_{2}=0$. The value of $r_{1}$ confirms the existence of singularity as one of the solutions of resonances must be $-1$. The second resonance confirms that the coefficient $a_{0}$ of the leading-order term is arbitrary, and it is another integration constant apart from $t_{0}$. Since the system contains two integration constants viz. $t_{0}$ and $a_{0}$, therefore, the system passes Painlev\'{e} test. However, to obtain some information on analytical solutions that are physically interesting, in what follows, we perform the last step of the
singularity analysis. Due to the presence of remainder terms arising  from the substitution of a leading term in \eqref{FRDEQ22}, the leading term is not a solution of the system. By inspecting the resonances, we can express the solution in the right Laurent expansion with step $\frac{1}{3}$ and so we have
\begin{equation}
a(\tau )=a_{0}\tau ^{\frac{2}{3}}+\sum_{i=1}^{+\infty }a_{i}\tau ^{\frac{2+i}{3}}\,.  \label{ps1}
\end{equation}%
At this point, we have determined the two free
parameters of the problem, coefficient $a_{0}$ and $t_0$ representing the position of the singularity. Hence, the consistency test is not necessary to conclude that the cosmological system possesses the Painlev\'{e} property. However, we perform the consistency test in order to determine the values of $a_{i}$($i=1,2,3,\ldots $). Performing the consistency test for a general $n$ is difficult as  comparison of terms is involved. Therefore, for the sake of completeness, we shall fix the value of $n$.

As an illustration, let us take $n=-1$ and we substitute the expression \eqref{ps1} in equations \eqref{FRDEQ11} and \eqref{FRDEQ22}. Then, we obtain $\rho_{m0}=\frac{4}{3} a_0^3$ and the non-zero coefficients are $a_{12k}$ ($k=1,2,..$) with $a_{12}=\frac{9}{320}\alpha a_{0}$,  $\frac{a_{24}}{a_{12}}=-\frac{33}{160}\alpha a_{0}$ etc. Similarly, on taking $n=-2$, we obtain $\rho_{m0}=\frac{4}{3} a_0^3$ and the non-zero coefficients are $a_{18k}$ ($k \in \mathbb{N}$) with $a_{18}=\frac{45}{3584}\alpha a_0$,  $\frac{a_{36}}{a_{18}}=-\frac{12195}{46592}  a_0$ etc. In both the cases we see that  many coefficients close to the dominant terms vanish and therefore the solution is approximated by the leading order term $a_0 \tau^\frac{2}{3}$ i.e. by a dust fluid-like solution.  Therefore,  near singularity, the solution \eqref{ps1} corresponds to a decelerated universe with $q=\frac{1}{2}$.  The presence of right Laurent expansion \eqref{ps1} implies that the matter dominated universe near singularity is
not a stable solution. This can also be confirmed from the saddle nature of
a matter dominated critical point $B$ (see Sec. \ref{sec:DSA}).

Indeed, in the Laurent expansion (\ref{ps1}) as we move far from the
singularity $\tau >0$, the  terms right from the leading-order
behaviour $\tau ^{\frac{2}{3}}$ dominates and describe the solution
of the differential equation. This type of property of the solution has been observed before in previous cosmological studies \cite{Paliathanasis:2016tch}. If the resonances were negative, then the analytic solutions would be expressed by a left Painlev\'{e} Series which indicates that the leading-order behaviour describes the attractor of a stable asymptotic solution.

One also may be interested in the investigation of the analytical solution in terms of the Hubble function. Since,  the scale factor $a(\tau)$ can be expressed in  terms of the Laurent expansion, the leading term of the solution of $H(\tau)$ near the putative singularity is given by $H(\tau)=p \tau^{-1}$ where $p$ is the leading exponent of $a(\tau)$. Near singularity, such solution describes the universe's era dominated by an ideal gas with $w_{\rm eff}=-1+\frac{2}{3p}$.  For the present case, we have $p=\frac{2}{3}$, which indeed describes the matter dominated era ($w_{\rm eff}=0$). On substituting $H(\tau )=\frac{2}{3}\tau^{-1}+m\tau ^{-1+r}$ in \eqref{FRDEQ22} and linearize around $m=0$, we obtain $r=-1$, which implies that the singularity is movable. Since the differential equation in $H$ is first order, we do not need to proceed with the analysis. However, for the sake of completeness,  we continue to perform the consistency test. For the test, we choose $n=-1$, then  the solution can be expressed as a Laurent expansion
\begin{equation}
H(\tau )=\frac{2}{3}\tau ^{-1}+\sum_{i=1}^{+\infty }H_{i}\tau ^{-1+i}\,, \label{psH1_nls_1}
\end{equation}
where the non-zero coefficients are $H_{4k}$ $(k=1,2, ..)$ with $H_4=\frac{9}{80}$, $H_8=-\frac{1269}{25600}$ etc. We note here that the only integration constant of equation \eqref{FRDEQ22} is $t_0$, which is  the location of the singularity. Thus, in the present case, the equation \eqref{FRDEQ22} possesses the Painlev\'e property and hence deemed to be integrable.

\subsubsection{$n>1$ case}

For the case $n>1$, we found that the leading order behavior is $\sigma=\frac{2n}{3}$ and $a_0$ is arbitrary i.e. $a(\tau)=a_0 \tau^\frac{2n}{3}$. Therefore, the dominant term comes from the geometric term $Q^n$ of the action. Depending on the value of $n$, we either have $\frac{2n}{3} \notin
\mathbb{N}$ or $\frac{2n}{3} \in \mathbb{N}$.

First let us assume the case where $\frac{2n}{3} \notin \mathbb{N}$. In this case, we obtain resonances $r_1=-1, 0$. Therefore, as in the case of $n<1$, the system passes the Painlev\'{e} test and the solution can be expressed in terms of a right Laurent expansion. As the step for the Laurent expansion depends on the value of $n$ therefore to find the series expansion we have to fix the value of $n$. For the sake of illustration, let us take $n=2$, then the right Laurent expansion is given by
\begin{equation}
a(\tau)=a_0 \tau^\frac{4}{3}+\sum_{i=1}^{+\infty} a_i \tau^{\frac{4+i}{3}}\,.
\label{ps2}
\end{equation}
Performing a consistency test, we get $\rho_{m0}=\frac{512}{3} \alpha a_0^3$ i.e. $\alpha>0$ for $\rho_{m0}>0$ and the non-zero coefficients $a_6=-\frac{a_0}{288 \alpha}$, $a_{12}=\frac{17 a_6}{2880 \alpha}$ etc.

On the other hand, for $\frac{2n}{3}\in \mathbb{N}$, say $\frac{2n}{3}=M$, to determine a movable singularity, we replace $a(\tau)$ by $b^{-1}(\tau)$ in the cosmological equations \eqref{FRDEQ11}-\eqref{FRDEQ22}. So, the dominant term is $b(\tau)=b_0 \tau^{-M}$ and resonances are $0, -1$. If we take $n=\frac{3}{2}$, the analytical solution of the scale factor is
\begin{equation}
a(\tau)^{-1}=b_0 \tau^{-1}+\sum_{i=1}^{+\infty} b_i \tau^{-1+i}\,.
\label{ps3}
\end{equation}
where $b_1=\frac{b_0}{12\sqrt{6} \alpha}$, $b_2=-\frac{b_0}{864 \alpha^2}$
etc., along with $\rho_{m0}=\frac{6\sqrt{6} \alpha}{b_0^3}$ which is positive for $\alpha>0$.

 In general, for $n>1$, the solution of the scale factor near the singularity approximated by the dominant term is given by the power-law solution $a(\tau)\propto \tau ^{\frac{2n}{3}}$, which corresponds to an effective fluid of equation of state parameter $w_{\mathrm{eff}}=\frac{1-n}{n}$. Therefore, the solution near singularity describes a decelerated universe for $1<n<\frac{3}{2}$, an accelerated universe for $n>\frac{3}{2}$, a Milne universe for $n=\frac{3}{2}$ and a cosmological constant as $n\rightarrow \infty $. We also note that the solution  run fast from the scaling behavior as this solution does not correspond to any asymptotic structure. For instance, by comparing the scaling solution $a(\tau)=a_0 \tau^{\frac{2n}{3}}$ with the effective equation of state parameter, one can find that $x=\frac{n-1}{2n}$ (with $y=0$), which is indeed not a critical point of the system \eqref{eq:xdys_ex}-\eqref{eq:ydys_ex}. 

 As in the previous case, here, we find the behavior of the Hubble function. For this case, the leading order behavior is $H(\tau)=\frac{2n}{3}\tau^{-1}$ and the resonance is $r=-1$ which confirms the existence of a movable singularity. For the consistency test, we choose $n=2$ and the Laurent expansion for $H$ is given by
\begin{equation}
H(\tau )=\frac{2n}{3}\tau ^{-1}+\sum_{i=1}^{+\infty }H_{i}\tau ^{-1+i}\,, \label{psH1_ngr_1}
\end{equation}
where the non-zero coefficients are $H_{2k}$ $(k=1,2, ..)$ with $H_2=-\frac{1}{144}$, $H_4=\frac{1}{17280}$ etc. Thus, in summary, we find that the present model possesses the Painlev\'{e} property in the presence of dust fluid  for any parameter $n$.

\subsection{Dust with radiation fluid}\label{ssec:dust_rad}

In this section, we shall consider the case where the
matter component includes dust with radiation. As in the previous case, we
focus our study on two different cases: $n<1$ and $n>1$.

\subsubsection{$n<1$ case}

Following a similar procedure as earlier, it follows from the equation \eqref{FRDEQSA}, that the leading term is $a(\tau)=a_0 \tau^{\frac{1}{2}}$. We
note here that $a_0$ is not arbitrary, but it is given by $a_0^4=\frac{4}{3}
\rho_{r0}$. Therefore, contrary to the case where the only dust is present,
the dominant behavior is radiation-like.

Substituting $a(\tau )=a_{0}\tau ^{\frac{1}{2}}+m\tau ^{\frac{1}{2}+r}$ in 
\eqref{FRDEQ22} and linearize around $m=0$, we obtained resonances $r_{1}=-1$
and $r_{2}=\frac{1}{2}$. The second resonance confirms that the coefficient $a_{0}$ is not arbitrary, instead, the coefficient $a_{1} $ is arbitrary (an integration constant). From the nature of resonances, the Painlev\'{e} series can be expressed
in a right Laurent expansion with step $\frac{1}{2}$, i.e.
\begin{equation}
a(\tau )=a_{0}\tau ^{\frac{1}{2}}+\sum_{i=1}^{+\infty }a_{i}\tau ^{\frac{1+i}{2}}\,.  \label{ps4}
\end{equation}%
To determine the values of $a_{i}$, we perform a consistency test by
substituting \eqref{ps4} in \eqref{FRDEQ11}-\eqref{FRDEQ22}. As in the case
of dust fluid only, we take $n=-1$ and obtain $a_{2}=-\frac{7}{8}\frac{a_{1}^{2}}{a_{0}}$, $a_{3}=\frac{5}{4}\frac{a_{1}^{3}}{a_{0}^2}$ etc., and $\rho_{r0}=\frac{3}{4}a_0^4$. Also, equation \eqref{FRDEQ11} yields $\rho_{m0}=\frac{9}{2}a_1 a_0^2$.  We have checked that $n=-2$ yields same values of coefficients $a_i$ as in the $n=-1$ case.

The connection with the critical points is similar to that with the case of dust, where in here  the leading order behavior is that of the radiation fluid, that is, point $A$. Therefore, the solution \eqref{ps4} corresponds to a decelerated universe with  $w_{\rm eff}=\frac{1}{3}$.  The presence of right Laurent expansion \eqref{ps4} implies the unstable nature
of a radiation dominated universe near a singularity, which accord with the
nature of radiation dominated critical point $A$ (see Sec. \ref{sec:DSA}).

\subsubsection{$n>1$ case}

For the case $n>1$, we found that the leading order behaviour is $\sigma=\frac{2n}{3}$. Therefore, the dominant term comes
from the geometric term $Q^n$ of the action. Similar to the case of dust
fluid case, we consider $n$ such that $\frac{2n}{3} \notin \mathbb{N}$ or $\frac{2n}{3} \in \mathbb{N}$. In both cases, we obtain resonances $0$ and $-1$. However, the corresponding Painlev\'{e} series fails to satisfy equations \eqref{FRDEQ11} and \eqref{FRDEQ22}. This result implies that the system does not possess the Painlev\'{e} property for $n>1$, which means we cannot solve the solution in Laurent expansion.

Hence, in the presence of dust fluid with radiation, the present model passes the Painlev\'{e} test when we work with $a(t)$  only for $n<1$.  We summarize the result of the singularity analysis performed in this section for the case of dust fluid and the dust fluid with radiation on Table \ref{tab:SAS}.  

Before we conclude this section, we remark that the system in the presence of dust with radiation fluid does not admit the Painlev\'e property when we work with the Hubble function $H(t)$. This is a common problem in the singularity analysis as the Painlev\'e property is coordinate dependent and hence depends on the equation or variables we apply, which is contrary to the symmetry analysis. For instance, the well-known integrable oscillator, in general, does not possess the Painlev\'e property, but one has to define coordinates to satisfy the property\cite{Paliathanasis_2016}. Similarly, the Starobinsky model of inflation admits the Painlev\'e property only for a specific choice of coordinates \cite{Paliathanasis:2017apr,Paliathanasis:2016tch}.  To further investigate the viability of the model, in the next section, we shall analyze the model's behavior at the linear perturbation level.

 \begin{table*}[!ht]
	\centering
	\renewcommand{\arraystretch}{1.8}
	\begin{tabular}{|*{6}{c|}}
		\hline
	{\bf	Cases}& {\bf Range of $n$}&~{\bf Painlev\'e Property} ~&~ {\bf Leading order of $a(\tau)$}~ &~{\bf Nature of solutions in $a(\tau)$}~ &{\bf  Dominant fluid} \\
			\hline\hline
				\multirow{2}*{Dust fluid} &$n<1$&Satisfied&$\tau^{\frac{2}{3}}$&Right Painlev\'e series&Dust\\
			\cline{2-6}
			&$n>1$&Satisfied&$\tau^{\frac{2n}{3}}$&Right Painlev\'e series&$F(Q)$\\		
				\hline
	\multirow{2}*{Dust fluid} &$n<1$& Satisfied&$\tau^{\frac{1}{2}}$&Right Painlev\'e series&Radiation\\			
\cline{2-6}
	 \multirow{1}*{with radiation}  &$n>1$& Inconclusive&Inconclusive&Undetermined&Undetermined\\
	 \hline
		\end{tabular}
	\caption{A summary on the result of  singularity analysis.}	\label{tab:SAS}	
\end{table*}

\section{Linear growth index}\label{sec:GI}
 In this section, we shall study the linear growth of dark matter
 fluctuations for the power-law form model  $F(Q)=\alpha Q^n$ in the matter dominated era (neglecting the radiation component, i.e. $\rho_{r}=0$). From \eqref{FRDEQ1}, one can find that the present matter density parameter  is
 \begin{equation}
 \Omega_{m0}=\frac{\rho_{m0}}{3H_0^2}=1-\frac{F(Q_0)}{6H_0^2}+(2F_Q)_{Q=Q_0}\,,
 \end{equation}
 where $Q_0=6H_0^2$ ($H_0$ is the Hubble constant) and so we have
 \begin{equation}
 \alpha=	(6H_0^2)^{1-n}\frac{1-\Omega_{m0}}{2n-1}\,.
 \end{equation}
 Now dividing equation \eqref{FRDEQ1} by $H_0^2$, we get
 \begin{equation}\label{E_def}
 E^2(a)=\frac{H^2}{H_0^2}=\Omega_{m0} a^{-3}+(1-\Omega_{m0}) E^{2n}\,.
 \end{equation}
 The  differential equation describing the evolution  of matter density perturbations $\delta$ defined by $\delta=\frac{\delta \rho_m}{\rho_m}$ at smaller scale  compared to the Hubble radius
 is given by \cite{Lue:2004rj,Linder:2004ng,Linder:2007hg}
 \begin{equation}\label{denpert}
 \ddot{\delta}+2H\nu \dot{\delta}-4\pi \mu \,\rho_m\delta=0\,.
 \end{equation}
 The quantities $\mu$ and $\nu$ are associated with the physics of DE and measure the deviation of GR's theory. For  DE models within the framework of GR, one has $\mu=\nu=1$. On the other hand, for various modified gravity theories, we have $\nu=1$ and $\mu \neq 1$. Further, if the matter component is allowed to couple with DE, we have $\nu \neq1$ and $\mu \neq 1$. Since, in our present model, there is no interaction between dark sectors, we have \cite{Jimenez:2019ovq}
 \begin{equation}\label{munufq}
 \nu=1\,,~~~~~~\mu=\frac{1}{1+F_Q}\,.
 \end{equation}
 In order to have a better picture on the evolution of growth of matter perturbation, it is convenient to consider the growth factor $f$ which is defined in terms of $\delta$ as \cite{peeblesbook} 
 \begin{equation}\label{grwth0}
 f\equiv \frac{d\ln\delta}{d\ln a}\,.
 \end{equation}
  In terms of $E$,  we can rewrite $\Omega_m$ as
 \begin{equation}\label{OmE}
 \Omega_m=\frac{\Omega_{m0}a^{-3}}{E^2(a)}\,.
 \end{equation}
 Differentiating \eqref{OmE} with respect to scale factor, we get
 \begin{equation}\label{diffOm}
 \frac{d\Omega_m}{da}=-3\frac{\Omega_{m}}{a} \left(1+\frac{2}{3} \frac{d \ln E}{d\ln a}\right)\,.
 \end{equation} 
 Using the definition \eqref{grwth0} and  equation \eqref{diffOm}, we can write an equation \eqref{denpert} as a first-order differential equation given by
 \begin{equation}\label{grwthfeq1}
 {d\; f\over d\ln a}+f^2+\bigg(2+\frac{d\ln E}{d\ln a}\bigg) f =\frac{3}{2}\frac{1}{1+F_Q}\Omega_m.
 \end{equation}
 On differentiating \eqref{E_def} with respect to $\ln a$  and using \eqref{OmE} we get
 \begin{eqnarray}
 \label{hubble0}
 {d \ln E\over d\ln a}=-\frac{3}{2}\frac{1-E^{2n-2}(1-\Omega_{m0})}{1-n E^{2n-2}(1-\Omega_{m0})}\,,
 \end{eqnarray}
 and so equation \eqref{grwthfeq1} becomes
 \begin{eqnarray}
 \label{eqg2}
 {d\; f\over d\ln a}+f^2 +f \left[2-{3\over2}\frac{1-E^{2n-2}(1-\Omega_{m0})}{1-nE^{2n-2}(1-\Omega_{m0})}\right]=\frac{3}{2}\frac{\Omega_{m}}{1-{n(1-\Omega_{m0})E^{2n-2}\over 2n-1}}\,.
 \end{eqnarray}
 As we are interested in the behavior of matter perturbations in the matter dominated era, we shall use the common parametrization of $f$  in terms of $\Omega_m$ given by  \cite{peeblesbook} 
 \begin{equation}\label{grwthf1}
 f=\Omega_m^\gamma (a)\,,
 \end{equation}
 where $\gamma$ is the growth index of matter perturbations.  It is worth mentioning that in the literature, there are various theoretical speculations on the functional form of the growth index.  In the present work, we consider the following phenomenologically interesting parametrization of $\gamma$ in terms of scale factor \cite{Basilakos:2012uu}:
 \begin{equation}\label{gam_par}
 \gamma(a) =\gamma_0 +\gamma_1 y(a)\,.
 \end{equation} 
 The above equation can be treated as a first-order Taylor  expansion of $\gamma$ around some cosmological function $y(a)$ with coefficients $\gamma_0, \gamma_1$. To determine the growth index's behavior, one has to specify the function $y(a)$.  In what follows, we shall investigate the growth index for two different parametrizations of $y(a)$. First, we consider the case where  $y(a)= \ln \Omega_{m}(a)$. Here, for $z\gg 1$ i.e. $\Omega_m(a) \to 1$, the growth index approach an asymptotic value $\gamma_\infty$ with $\gamma_\infty \approx \gamma_0$ which is redshift independent. Such a form of $\gamma$ is also called the constant growth index. We consider another form  of $y(a)$  given by $y(a)=1-a$. In this case, the growth index is redshift-dependent, and therefore, it is known as a time-varying growth index. We note that the growth index's redshift-dependent form can provide a more accurate approximation for the growth rate factor compared to a constant parametrization. In what follows, we shall discuss the constant growth index's behavior and the time-varying growth index separately.

 \subsection{Constant growth index}\label{sec:CGI}
 In this subsection, we consider the simplest form of the growth index known as the asymptotic or constant growth index. For finding the value of the asymptotic growth index, we use an analytical approach developed in \cite{Steigerwald:2014ava}. Based on an analytical method, for high redshift $z \gg 1$ i.e. $\Omega_{m} \to 1$, the asymptotic growth index $\gamma_\infty$ is  given by
 \begin{equation}
 \gamma_\infty=\frac{3(M_0+M_1)-2(H_1+N_1)}{2-4H_1+3 M_0}\,,
 \end{equation}
 where
 \begin{gather}
 M_0=\mu\Big|_{\Omega_{m}=1}, ~~~M_1=\frac{d\mu}{d\ln \Omega_{m}}\Big|_{\Omega_{m}=1},~~~~N_1=\frac{d\nu}{d\ln \Omega_{m}}\Big|_{\Omega_{m}=1}, ~~~~H_1=\frac{d\left(\frac{d\ln E}{d\ln a}\right)}{d\ln \Omega_{m}}\Big|_{\Omega_{m}=1}\,.
 \end{gather}
 We note here that $$\frac{d\mu}{d\ln \Omega_{m}}=\frac{\Omega_{m}\,n(1-2n)}{\Omega_m\, n+n-1}.$$
 After some algebraic calculations, we obtain
 \begin{gather}
 M_0=1, ~~~~~M_1=\left\lbrace
 \begin{array}{cc}
 0 & \text{if}~n=\frac{1}{2}\\[1ex]
 \frac{n}{1-2n} & \text{if}~n \neq \frac{1}{2}
 \end{array}
 \right. ,~~~~~N_1=0, ~~~~~H_1=\frac{3(n-1)}{2}\,.
 \end{gather}
 Therefore, the value of the asymptotic growth index is given by
 \begin{equation}\label{ginffQ}
 \gamma_\infty=\left\lbrace
 \begin{array}{cc}
 \frac{9}{16} & \text{if}~n=\frac{1}{2}\,,\\[1ex]
 \frac{6(n-1)^2}{(2n-1)(-11+6n)} & \text{if}~n \neq \frac{1}{2} \,.
 \end{array}
 \right.
 \end{equation}
 \begin{figure}[tbp]
 	\centering
 	\subfigure[]{
 		\includegraphics[width=7cm, height=7cm]{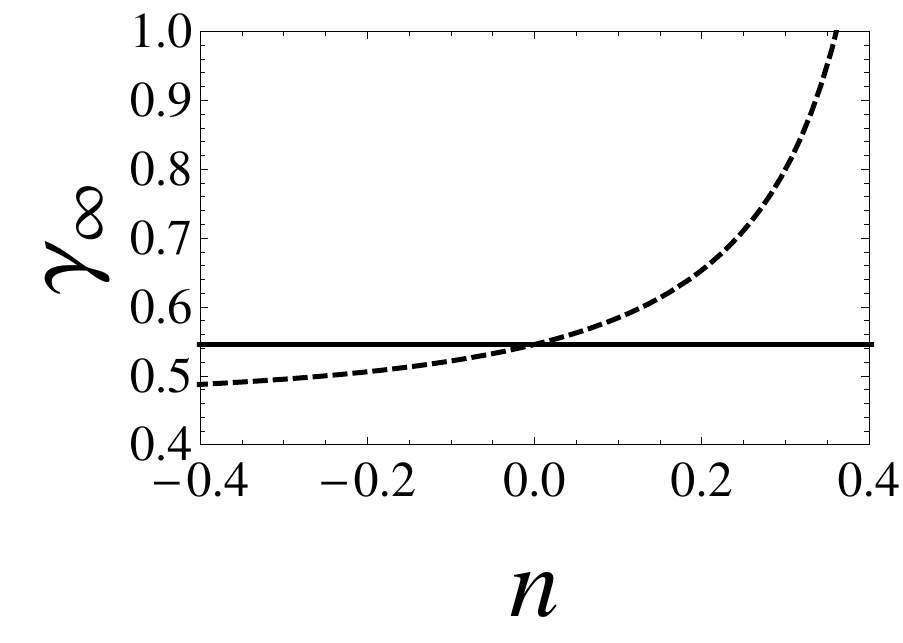} \label{fig:g_inf_1}}
 	\qquad
 	\subfigure[]{
 		\includegraphics[width=7cm, height=7cm]{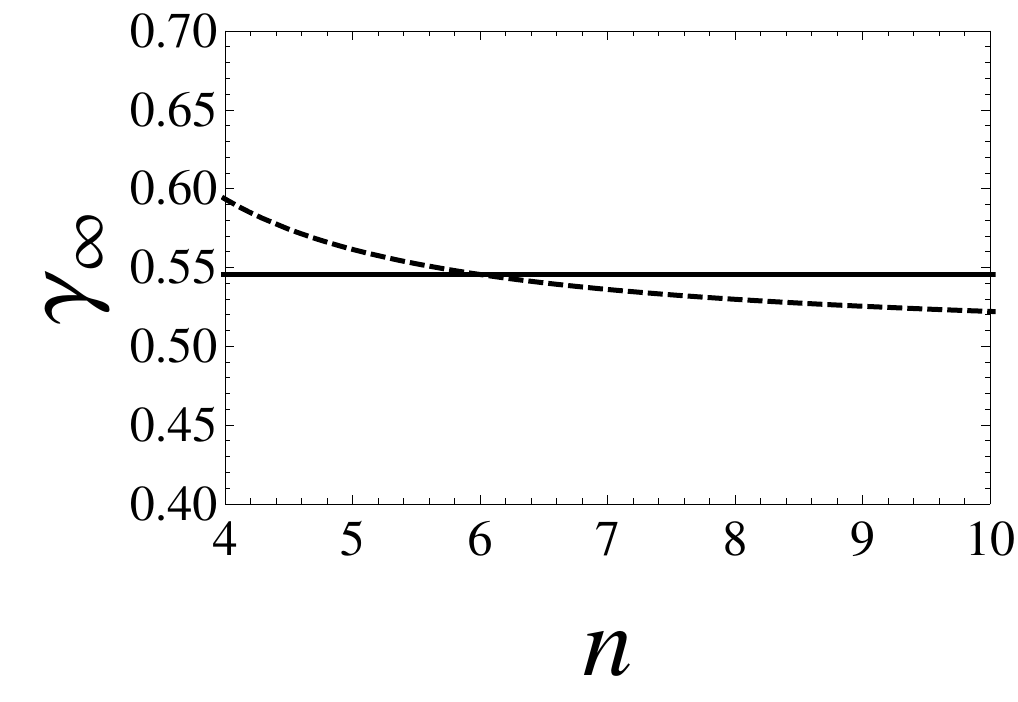} \label{fig:g_inf_2} }
 	\caption{The evolution of $\gamma_\infty$ for the power law model $F(Q)=\alpha Q^n$ as a function of parameter $n$ (in dashed curve). The solid line $\gamma_\infty=\frac{6}{11}$ corresponds to the $\Lambda$CDM model i.e. $n=0$.}
 	\label{fig:g_inf}
 \end{figure}
  Indeed, for $n=0$, one recover the standard value $\frac{6}{11}$ for the $\Lambda$CDM model. We note that the value of $\gamma_\infty$ is undefined for $n= \frac{11}{6}$. Interestingly,  for $n=\frac{1}{2}$, the value of $\gamma_\infty=\frac{9}{16}$ which coincide with that of the Finsler-Randers cosmological model \cite{Papagiannopoulos:2017whb}. We found some differences in the value of the growth index from the GR-based model, i.e. the $\Lambda$CDM, while background cosmology for   $n=\frac{1}{2}$ is the same as that of GR for any matter content \cite{Jimenez:2019ovq}.  In Fig. \ref{fig:g_inf}, we plot the asymptotic growth index as a function of a parameter $n$ and compare it with that of the $\Lambda$CDM model. The plot shows that the value of the growth index for the power-law model is greater than that of $\Lambda$CDM only for  $0<n<\frac{1}{2}$ or $\frac{11}{6}<n<6$. As mentioned earlier, the varying growth index contains more information about the growth of structures described by gravity's underlying theory; therefore, we now proceed to analyze a time-varying growth index.
  
 \subsection{Varying growth index}\label{sec:VGI}
 For the redshift-dependent growth index, we consider the parameterization introduced in \cite{Polarski:2007rr} which is expressed as a Taylor expansion around the present time i.e. $a(z)=1$ as:
 \begin{equation}\label{vargind}
 \gamma(a)=\gamma_0+\gamma_1 (1-a)\,,
 \end{equation}
where $\gamma_0, \gamma_1$ are the coefficients.  The parametrization \eqref{vargind}  is commonly used in the literature to approximate the growth rate of matter perturbation. It provides a very good approximation to the $\Lambda$CDM and DGP models. Here, we shall  analyze  it for the power law form of $f(Q)$  gravity. On substituting $f(a)$ on \eqref{grwthfeq1} and using \eqref{diffOm}, yields
 \begin{equation}\label{dgama_da}
 a \ln (\Omega_m) \frac{d \gamma}{d a}+\Omega^\gamma_m-3\Big(\gamma-\frac{1}{2}\Big) \Big(1+\frac{2}{3} \frac{d \ln E}{d \ln a}\Big)+\frac{1}{2}-\frac{3}{2}\,\mu\, \Omega^{1-\gamma}_m=0\,.
 \end{equation}
 At the present time which corresponds to $a=1$ i.e. redshift $z=0$, the above equation becomes
 \begin{eqnarray}\label{dgama_da_1}
 \frac{d \gamma}{d a}\big|_{a=1}\,\ln \Omega_{m0}+ \Omega^{\gamma(1)}_{0}-3\Big(\gamma(1)-\frac{1}{2}\Big)\,\Big(1+\frac{2}{3}\,\frac{d \ln E}{d \ln a}\Big)_{a=1}+\frac{1}{2}-\frac{3}{2}\,\mu_0\,\Omega^{1-\gamma(1)}_{m0}=0\,,
 \end{eqnarray}
 where $\mu_0=\mu\big|_{a=1}=\frac{2n-1}{n\, \Omega_{m0}+n-1}$ and $\frac{d \ln E}{d \ln a}\big|_{a=1}=-\frac{3}{2}\frac{\Omega_{m0}}{1-n (1-\Omega_{m0})}$.
 By employing the  parametrization \eqref{vargind} in equation \eqref{dgama_da_1}, we obtain $\gamma_1$ in terms of $\gamma_0$ as
 \begin{eqnarray}\label{g0g1_rel}
 \gamma_1=\frac{1}{\ln (\Omega_{m0})} \Big[\Omega_{m0}^{\gamma_0}-3\left(\gamma_0-\frac{1}{2}\right) \left(\frac{(\Omega_{m0}-1)(n-1)}{n(\Omega_{m0}-1)+1}\right)+\frac{1}{2}-\frac{3}{2} \mu_0 \Omega_{m0}^{1-\gamma_0}\Big]\,.
 \end{eqnarray}
 Again for large redshift $z\gg 1$ (or $a(z) \to 0$), we have $\gamma=\gamma_\infty$ and so from \eqref{vargind}, it can be seen that $\gamma_\infty \simeq \gamma_0+\gamma_1$. Therefore, using the expression of $\gamma_\infty$ from equation \eqref{ginffQ}, one can find the expressions of $\gamma_0$ and $\gamma_1$ in terms of $\Omega_{m0}$ and $n$.
 
 \begin{figure}[tbp]
 	\centering
 	\subfigure[]{
 		\includegraphics[width=7cm, height=7cm]{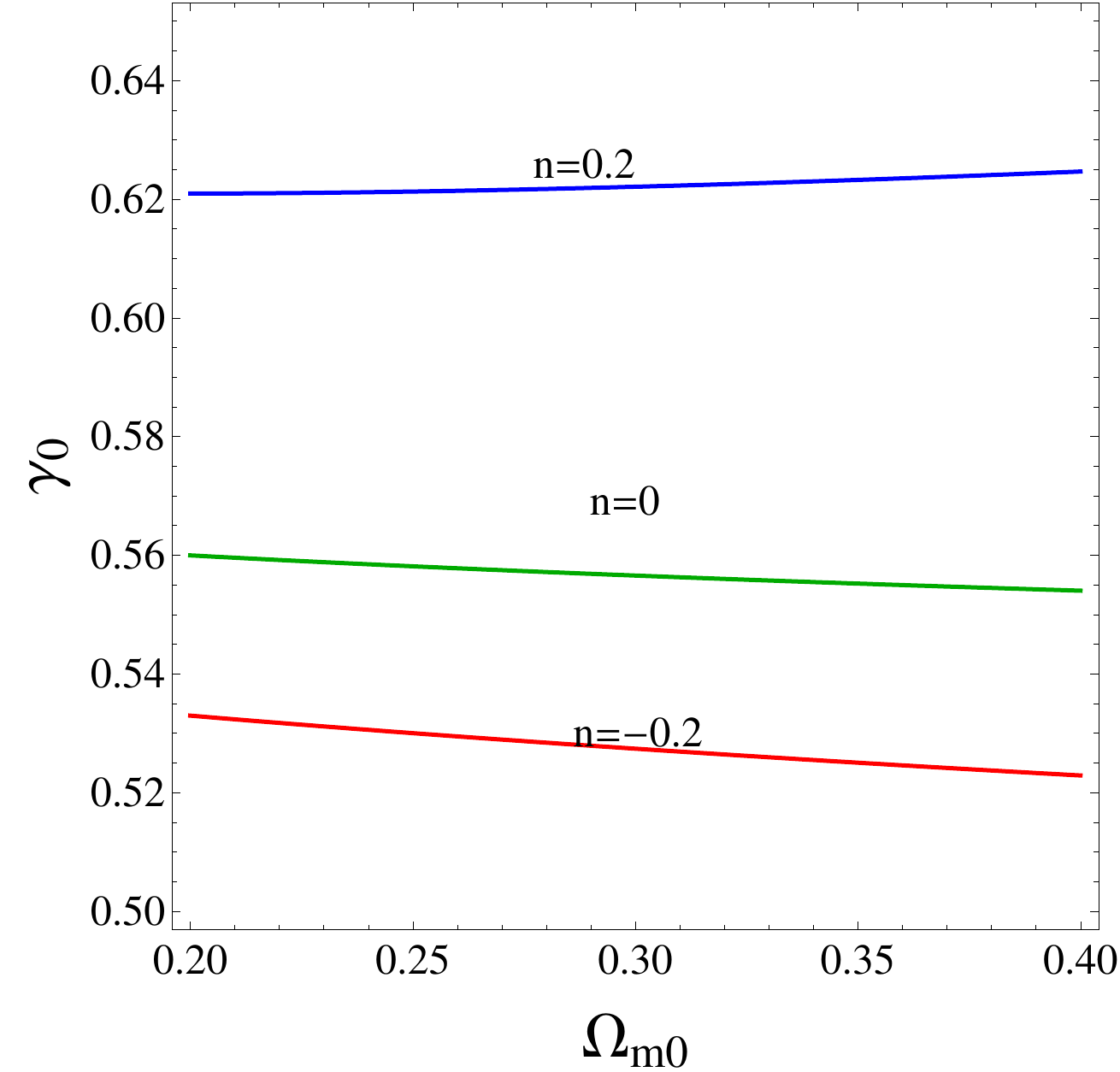} \label{fig:g0}}
 	\qquad
 	\subfigure[]{
 		\includegraphics[width=7cm, height=7cm]{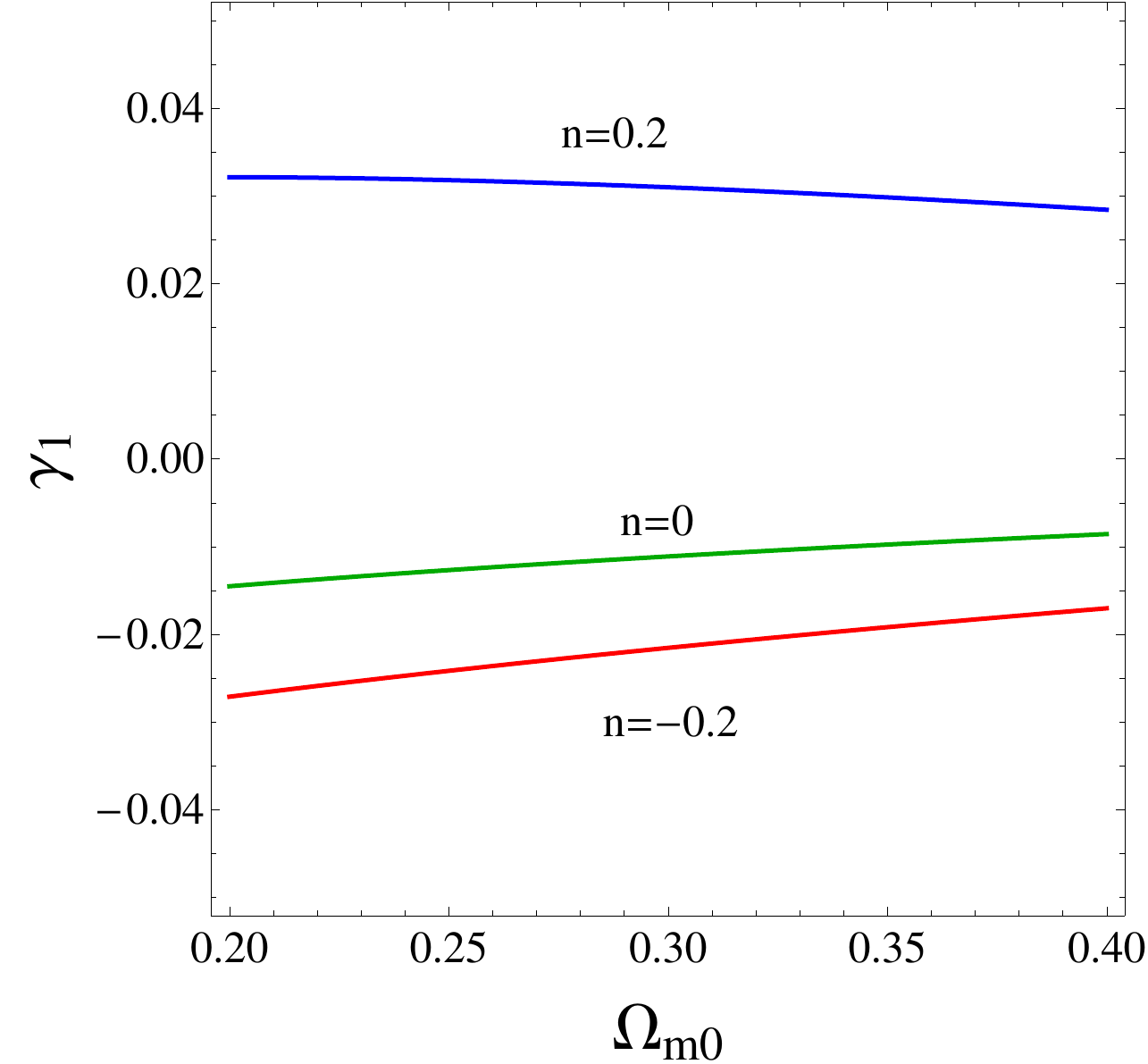} \label{fig:g1} }
 	\caption{Evolution of $\gamma_0$ and $\gamma_1$ against $\Omega_{m0}$ for $n=-0.2, ~0.2$ and $n=0$ which corresponds to the $\Lambda$CDM model.}
 	\label{fig:g0g1}
 \end{figure}
 
 \begin{figure}[h!]
 	\centering
 	\includegraphics[width=10cm, height=7cm]{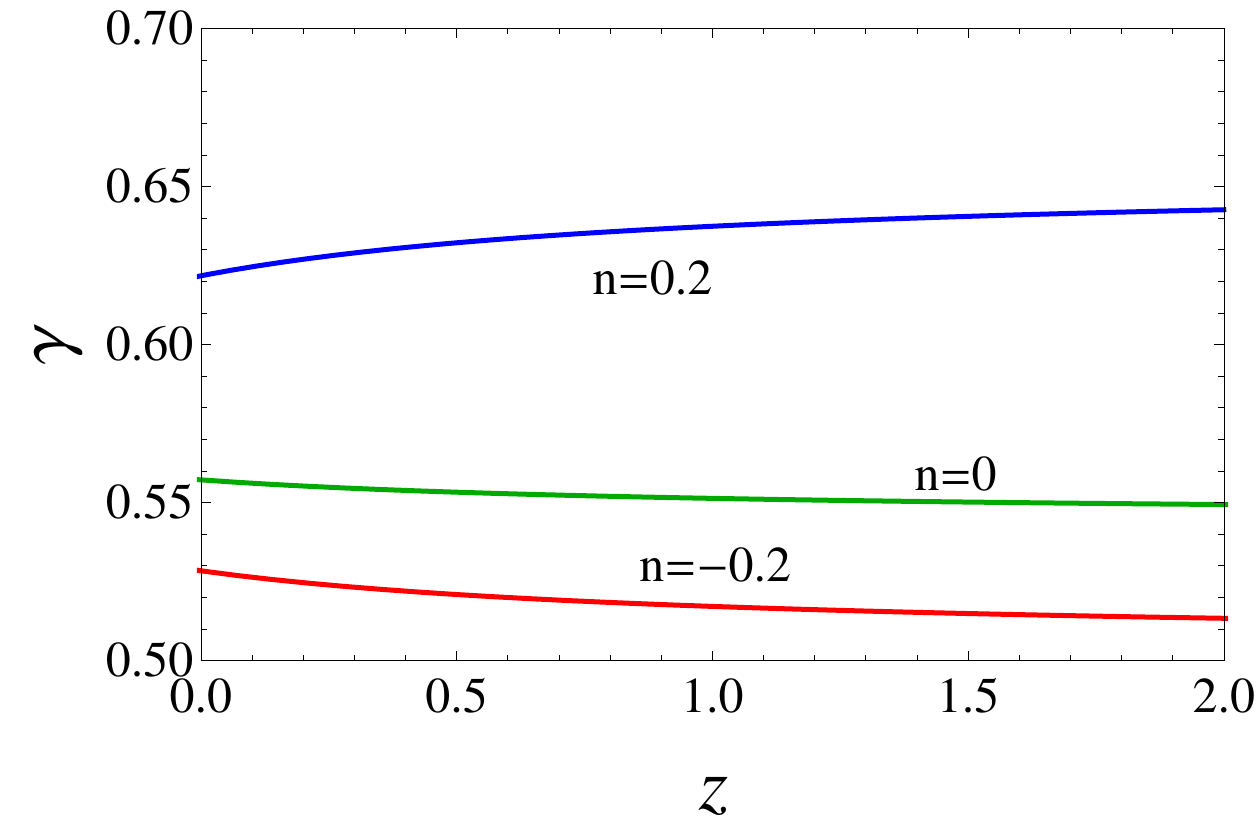}
 	\caption{The evolution of growth index $\gamma$ against redshift $z$ for $n=-0.2,~0.2$ and for the $\Lambda$CDM model i.e. $n=0$ with $\Omega_{m0}=0.28$\,.}
 	\label{fig:gindex}
 \end{figure}
 
 In Fig. \ref{fig:g0g1}, we plot the variation of quantities $\gamma_0$ and $\gamma_1$ for $0.2\leq \Omega_{m0} \leq 0.4$.  We find that   the values of $\gamma_0$ and $\gamma_1$ are larger (or smaller) than those in the $\Lambda$CDM model (i.e. $n=0$) for $n>0$ (or $n<0$). This result implies that the  effective gravity of the $f(Q)$ theory is weaker for $n>0$ and stronger for $n<0$ in comparison to GR. Such features provide distinct signatures for the $f(Q)$ theory. It can be also noted that for various  GR based models, we have $\vert \gamma_1 (z=0) \vert \lesssim 0.02$ and can be greater for models beyond GR framework \cite{Gannouji:2008jr}. Hence in principle, one can discriminate the $f(Q)$ gravity from Einstein gravity through the values of $\gamma_0, \gamma_1$.

 Finally, in Fig. \ref{fig:gindex}, we plot the growth index's evolution for positive and negative values of $n$. We can see that throughout the evolution, the growth index is smaller in comparison to that of $\Lambda$CDM for $n<0$ and larger for $n>0$. Such behavior of the growth index can be related to the nature of the effective Newton's gravitational constant (see equation \eqref{munufq}). As the growth index's evolution depends on the value of parameter $n$,  one needs to perform constraint on $\gamma_0, \gamma_1$ concerning various observational growth rate data. This exercise will restrict the possible values of $n$ favorable with the structure formation data.

\section{Conclusion}\label{sec:conc}
We studied the recently modified gravity theory's cosmological behavior known as the $f(Q)$ theory in the present work.  More accurately, we focussed the analysis on the power-law form, i.e., $f(Q)=Q+\alpha Q^n$. While the theory modifies high curvature regime for $n>1$, it modifies low curvature regime for $n<1$. Therefore, the theory applies to the early Universe for $n>1$ and to the late time Universe for $n<1$. It is worth mentioning that for $n=\frac{1}{2}$, the background evolution is the same as that of GR. However, Ref. \cite{Frusciante:2021sio} reported some interesting observational signatures distinct from that of the $\Lambda$CDM model at the perturbation level. Therefore, in this work, we have investigated the behavior of the power-law model of $f(Q)$ theory at the background level and perturbation level.

At the background level, we perform a dynamical system analysis and singularity analysis of the model.  From the dynamical system analysis performed in Sec. \ref{sec:DSA}, we see that for $|n|<1$, the model is cosmologically viable, exhibiting a cosmological sequence: radiation era $\to$ dark matter era $\to$ DE era. Therefore, for such a case, the overall background evolution is the same as that of the $\Lambda$CDM. It is worth mentioning that the DE 
behavior is solely due to the geometry of the theory without the need to introduce an exotic component.

A dynamical system usually contains model parameters and initial conditions, which we require to fine-tune for performing numerical analysis. In such a scenario, one may not understand the system's general properties defined by the theory. Therefore, finding the analytical solutions for given differential equations is crucial to understand their properties further. In the present work, we determine the analytical solutions of the scale factor and the Hubble function by employing the singularity analysis techniques in Sec. \ref{sec:SAM}. We followed the ARS algorithm, which allows us to determine whether the differential equation possesses the Painlev\'{e} property. Further, we compared the Laurent expansion with the nature of critical points obtained from the dynamical system analysis. More precisely, we performed the singularity analysis for two cases: (a) dust fluid only in subsection \ref{ssec:dust} (b) dust fluid along with radiation in subsection \ref{ssec:dust_rad}.

In the case of dust only, the system passes the Painlev\'{e} test for both $n<1$ and $n>1$. For $n<1$, the solution around singularity is approximated by a dust fluid-like solution, and it is expressed as a right Painlev\'{e} series. The nature of the series also confirmed the saddle behavior of a matter dominated critical point $B$ of the system \eqref{eq:xdys_ex}-\eqref{eq:ydys_ex}. If $n>1$, the solution near singularity corresponds to a scaling solution due to the geometric term $Q^n$. It describes decelerated Universe for $1<n<\frac{3}{2}$, an accelerated Universe for $n>\frac{3}{2}$,  a Milne universe for $n=\frac{3}{2}$ and a cosmological constant as $n\rightarrow \infty $. However, we note that the solution does not correspond to any asymptotic structure of a dynamical system.

In the presence of dust and radiation, the system passes the Painlev\'{e} test only for  $n<1$.  In such a case, the solution around singularity is approximated by a radiation-like solution, and it is expressed as a right Painlev\'{e} series.  As before, the nature of the series also confirmed the unstable nature of radiation dominated critical point $A$ of the system.  Therefore, the singularity analysis results complement the dynamical system analysis performed in Sec. \ref{sec:DSA}.  Failure of the Painlev\'e test in the presence of radiation and matter for $n>1$ suggests that the system is likely to be non-integrable. The non-integrability of the system may lead to the onset of chaotic behavior. Therefore, we believe that the singularity analysis allows us to understand the onset of chaos in gravity theories. Most importantly, the present work contributes to the subject of integrability of gravitational field equations in the context of
cosmology. 

The singularity analysis  for the power-law form of $f(R)$ gravity i.e. $f(R)=R+qR^n$ has been performed   in \cite{Paliathanasis:2016tch,Paliathanasis:2017apr} for $n>1$ and $n\neq  \frac{5}{4}$. The analysis is carried out in the absence of radiation and for $n>1$ only. However, no such analysis have been reported for $n<1$ or in the presence of radiation for $f(R)$ gravity. By comparing the results of $f(R)$ gravity with that obtained in the present work, we find that both theories provide similar dominant power-law solutions except $n=2$.  Hence, both $f(R)$ and $f(Q)$ theories resemble near a movable singularity for $n>1$.

The knowledge about the integrability of the field equations is crucial for the existence and determining real solutions. We note that apart from the singularity analysis, the symmetry analysis which is coordinate independent  is another method that allows us to extract information if the system is integrable.  Therefore, in the future we aim to use both approaches in other theories of gravity.

Apart from the background analysis, we studied the growth rate of matter perturbations within the sub-horizon scale. In particular, we analyze the growth rate index's nature for two different choices: the asymptotic value of growth index and varying form of growth index. We find that the value of the asymptotic growth index depends on parameter $n$. For $n=0$, we recover the standard value of the $\Lambda$CDM which is $\frac{6}{11}$.  For $n=\frac{1}{2}$, it is well-known that at the background level, the dynamics are the same as that of GR \cite{Jimenez:2019ovq}. Such a result is further confirmed by the dynamical system analysis performed in Sec. \ref{sec:DSA}. However, we find that the value of the asymptotic growth index for $n=\frac{1}{2}$  is $\frac{9}{16}$ which interestingly deviates from that of the $\Lambda$CDM model.  More specifically, the value of growth index for the present model is greater than that of $\Lambda$CDM when  $0<n<\frac{1}{2}$ or $\frac{11}{6}<n<6$, otherwise it is smaller than that of $\Lambda$CDM (see Fig. \ref{fig:g_inf}). Finally, we generalized the analysis by considering the growth index parametrization, which varies with redshift introduced in \cite{Polarski:2007rr}. The investigation reveals that throughout the evolution, the varying growth index is smaller to that of the $\Lambda$CDM for $n<0$ and larger for $n>0$, which can be related to the nature of gravity  (see parameter $\mu$ from equation \eqref{munufq}). The results obtained here also confirmed the possible peculiar or measurable signature at the linear regime of cosmic perturbations investigated in \cite{Frusciante:2021sio}. Hence, we require further analysis to test the theory's viability with upcoming precise observational data at the perturbation level.

\begin{acknowledgments}
	
	 JD was supported by the Core Research Grant of SERB, Department of Science and Technology India (File No. CRG $\slash 2018 \slash 001035$) and the Associate program of IUCAA. This work is based on the research supported in part by the National Research Foundation of South Africa (Grant Numbers 131604). We are thankful to the referee for the comments which helped us to improve the manuscript.
\end{acknowledgments}

\bibliography{fqpainleve}

\begin{thebibliography}{71}%
\makeatletter
\providecommand \@ifxundefined [1]{%
 \@ifx{#1\undefined}
}%
\providecommand \@ifnum [1]{%
 \ifnum #1\expandafter \@firstoftwo
 \else \expandafter \@secondoftwo
 \fi
}%
\providecommand \@ifx [1]{%
 \ifx #1\expandafter \@firstoftwo
 \else \expandafter \@secondoftwo
 \fi
}%
\providecommand \natexlab [1]{#1}%
\providecommand \enquote  [1]{``#1''}%
\providecommand \bibnamefont  [1]{#1}%
\providecommand \bibfnamefont [1]{#1}%
\providecommand \citenamefont [1]{#1}%
\providecommand \href@noop [0]{\@secondoftwo}%
\providecommand \href [0]{\begingroup \@sanitize@url \@href}%
\providecommand \@href[1]{\@@startlink{#1}\@@href}%
\providecommand \@@href[1]{\endgroup#1\@@endlink}%
\providecommand \@sanitize@url [0]{\catcode `\\12\catcode `\$12\catcode
  `\&12\catcode `\#12\catcode `\^12\catcode `\_12\catcode `\%12\relax}%
\providecommand \@@startlink[1]{}%
\providecommand \@@endlink[0]{}%
\providecommand \url  [0]{\begingroup\@sanitize@url \@url }%
\providecommand \@url [1]{\endgroup\@href {#1}{\urlprefix }}%
\providecommand \urlprefix  [0]{URL }%
\providecommand \Eprint [0]{\href }%
\providecommand \doibase [0]{http://dx.doi.org/}%
\providecommand \selectlanguage [0]{\@gobble}%
\providecommand \bibinfo  [0]{\@secondoftwo}%
\providecommand \bibfield  [0]{\@secondoftwo}%
\providecommand \translation [1]{[#1]}%
\providecommand \BibitemOpen [0]{}%
\providecommand \bibitemStop [0]{}%
\providecommand \bibitemNoStop [0]{.\EOS\space}%
\providecommand \EOS [0]{\spacefactor3000\relax}%
\providecommand \BibitemShut  [1]{\csname bibitem#1\endcsname}%
\let\auto@bib@innerbib\@empty
%</preamble>
\bibitem [{\citenamefont {Jim\'enez}\ \emph {et~al.}(2019)\citenamefont
  {Jim\'enez}, \citenamefont {Heisenberg},\ and\ \citenamefont
  {Koivisto}}]{BeltranJimenez:2019tjy}%
  \BibitemOpen
  \bibfield  {author} {\bibinfo {author} {\bibfnamefont {Jose~Beltr\'an}\
  \bibnamefont {Jim\'enez}}, \bibinfo {author} {\bibfnamefont {Lavinia}\
  \bibnamefont {Heisenberg}}, \ and\ \bibinfo {author} {\bibfnamefont
  {Tomi~S.}\ \bibnamefont {Koivisto}},\ }\bibfield  {title} {\enquote {\bibinfo
  {title} {{The Geometrical Trinity of Gravity}},}\ }\href {\doibase
  10.3390/universe5070173} {\bibfield  {journal} {\bibinfo  {journal}
  {Universe}\ }\textbf {\bibinfo {volume} {5}},\ \bibinfo {pages} {173}
  (\bibinfo {year} {2019})},\ \Eprint {http://arxiv.org/abs/1903.06830}
  {arXiv:1903.06830 [hep-th]} \BibitemShut {NoStop}%
\bibitem [{\citenamefont {Aldrovandi}\ and\ \citenamefont
  {Pereira}(2013)}]{aldro}%
  \BibitemOpen
  \bibfield  {author} {\bibinfo {author} {\bibfnamefont {Ruben}\ \bibnamefont
  {Aldrovandi}}\ and\ \bibinfo {author} {\bibfnamefont {Jos\'e~Geraldo}\
  \bibnamefont {Pereira}},\ }\href {\doibase
  https://doi.org/10.1007/978-94-007-5143-9} {\emph {\bibinfo {title}
  {{Teleparallel Gravity}}}},\ Vol.\ \bibinfo {volume} {173}\ (\bibinfo
  {publisher} {Springer, Dordrecht},\ \bibinfo {year} {2013})\BibitemShut
  {NoStop}%
\bibitem [{\citenamefont {Nester}\ and\ \citenamefont
  {Yo}(1999)}]{Nester:1998mp}%
  \BibitemOpen
  \bibfield  {author} {\bibinfo {author} {\bibfnamefont {James~M.}\
  \bibnamefont {Nester}}\ and\ \bibinfo {author} {\bibfnamefont {Hwei-Jang}\
  \bibnamefont {Yo}},\ }\bibfield  {title} {\enquote {\bibinfo {title}
  {{Symmetric teleparallel general relativity}},}\ }\href@noop {} {\bibfield
  {journal} {\bibinfo  {journal} {Chin. J. Phys.}\ }\textbf {\bibinfo {volume}
  {37}},\ \bibinfo {pages} {113} (\bibinfo {year} {1999})},\ \Eprint
  {http://arxiv.org/abs/gr-qc/9809049} {arXiv:gr-qc/9809049} \BibitemShut
  {NoStop}%
\bibitem [{\citenamefont {Altschul}\ \emph {et~al.}(2015)\citenamefont
  {Altschul} \emph {et~al.}}]{Altschul:2014lua}%
  \BibitemOpen
  \bibfield  {author} {\bibinfo {author} {\bibfnamefont {Brett}\ \bibnamefont
  {Altschul}} \emph {et~al.},\ }\bibfield  {title} {\enquote {\bibinfo {title}
  {{Quantum tests of the Einstein Equivalence Principle with the
  STE\textendash{}QUEST space mission}},}\ }\href {\doibase
  10.1016/j.asr.2014.07.014} {\bibfield  {journal} {\bibinfo  {journal} {Adv.
  Space Res.}\ }\textbf {\bibinfo {volume} {55}},\ \bibinfo {pages} {501--524}
  (\bibinfo {year} {2015})},\ \Eprint {http://arxiv.org/abs/1404.4307}
  {arXiv:1404.4307 [gr-qc]} \BibitemShut {NoStop}%
\bibitem [{\citenamefont {Beltr\'an~Jim\'enez}\ \emph
  {et~al.}(2018)\citenamefont {Beltr\'an~Jim\'enez}, \citenamefont
  {Heisenberg},\ and\ \citenamefont {Koivisto}}]{BeltranJimenez:2017tkd}%
  \BibitemOpen
  \bibfield  {author} {\bibinfo {author} {\bibfnamefont {Jose}\ \bibnamefont
  {Beltr\'an~Jim\'enez}}, \bibinfo {author} {\bibfnamefont {Lavinia}\
  \bibnamefont {Heisenberg}}, \ and\ \bibinfo {author} {\bibfnamefont {Tomi}\
  \bibnamefont {Koivisto}},\ }\bibfield  {title} {\enquote {\bibinfo {title}
  {{Coincident General Relativity}},}\ }\href {\doibase
  10.1103/PhysRevD.98.044048} {\bibfield  {journal} {\bibinfo  {journal} {Phys.
  Rev. D}\ }\textbf {\bibinfo {volume} {98}},\ \bibinfo {pages} {044048}
  (\bibinfo {year} {2018})},\ \Eprint {http://arxiv.org/abs/1710.03116}
  {arXiv:1710.03116 [gr-qc]} \BibitemShut {NoStop}%
\bibitem [{\citenamefont {Beltr\'an~Jim\'enez}\ \emph
  {et~al.}(2020)\citenamefont {Beltr\'an~Jim\'enez}, \citenamefont
  {Heisenberg}, \citenamefont {Koivisto},\ and\ \citenamefont
  {Pekar}}]{Jimenez:2019ovq}%
  \BibitemOpen
  \bibfield  {author} {\bibinfo {author} {\bibfnamefont {Jose}\ \bibnamefont
  {Beltr\'an~Jim\'enez}}, \bibinfo {author} {\bibfnamefont {Lavinia}\
  \bibnamefont {Heisenberg}}, \bibinfo {author} {\bibfnamefont
  {Tomi~Sebastian}\ \bibnamefont {Koivisto}}, \ and\ \bibinfo {author}
  {\bibfnamefont {Simon}\ \bibnamefont {Pekar}},\ }\bibfield  {title} {\enquote
  {\bibinfo {title} {{Cosmology in $f(Q)$ geometry}},}\ }\href {\doibase
  10.1103/PhysRevD.101.103507} {\bibfield  {journal} {\bibinfo  {journal}
  {Phys. Rev. D}\ }\textbf {\bibinfo {volume} {101}},\ \bibinfo {pages}
  {103507} (\bibinfo {year} {2020})},\ \Eprint
  {http://arxiv.org/abs/1906.10027} {arXiv:1906.10027 [gr-qc]} \BibitemShut
  {NoStop}%
\bibitem [{\citenamefont {Sotiriou}\ and\ \citenamefont
  {Faraoni}(2010)}]{Sotiriou:2008rp}%
  \BibitemOpen
  \bibfield  {author} {\bibinfo {author} {\bibfnamefont {Thomas~P.}\
  \bibnamefont {Sotiriou}}\ and\ \bibinfo {author} {\bibfnamefont {Valerio}\
  \bibnamefont {Faraoni}},\ }\bibfield  {title} {\enquote {\bibinfo {title}
  {{f(R) Theories Of Gravity}},}\ }\href {\doibase 10.1103/RevModPhys.82.451}
  {\bibfield  {journal} {\bibinfo  {journal} {Rev. Mod. Phys.}\ }\textbf
  {\bibinfo {volume} {82}},\ \bibinfo {pages} {451--497} (\bibinfo {year}
  {2010})},\ \Eprint {http://arxiv.org/abs/0805.1726} {arXiv:0805.1726 [gr-qc]}
  \BibitemShut {NoStop}%
\bibitem [{\citenamefont {Di~Valentino}\ \emph {et~al.}(2021)\citenamefont
  {Di~Valentino}, \citenamefont {Mena}, \citenamefont {Pan}, \citenamefont
  {Visinelli}, \citenamefont {Yang}, \citenamefont {Melchiorri}, \citenamefont
  {Mota}, \citenamefont {Riess},\ and\ \citenamefont {Silk}}]{re1}%
  \BibitemOpen
  \bibfield  {author} {\bibinfo {author} {\bibfnamefont {Eleonora}\
  \bibnamefont {Di~Valentino}}, \bibinfo {author} {\bibfnamefont {Olga}\
  \bibnamefont {Mena}}, \bibinfo {author} {\bibfnamefont {Supriya}\
  \bibnamefont {Pan}}, \bibinfo {author} {\bibfnamefont {Luca}\ \bibnamefont
  {Visinelli}}, \bibinfo {author} {\bibfnamefont {Weiqiang}\ \bibnamefont
  {Yang}}, \bibinfo {author} {\bibfnamefont {Alessandro}\ \bibnamefont
  {Melchiorri}}, \bibinfo {author} {\bibfnamefont {David~F.}\ \bibnamefont
  {Mota}}, \bibinfo {author} {\bibfnamefont {Adam~G.}\ \bibnamefont {Riess}}, \
  and\ \bibinfo {author} {\bibfnamefont {Joseph}\ \bibnamefont {Silk}},\
  }\bibfield  {title} {\enquote {\bibinfo {title} {{In the Realm of the Hubble
  tension $-$ a Review of Solutions}},}\ }\href@noop {} {\  (\bibinfo {year}
  {2021})},\ \Eprint {http://arxiv.org/abs/2103.01183} {arXiv:2103.01183
  [astro-ph.CO]} \BibitemShut {NoStop}%
\bibitem [{\citenamefont {Yang}\ \emph {et~al.}(2021)\citenamefont {Yang},
  \citenamefont {Pan}, \citenamefont {Di~Valentino}, \citenamefont {Mena},\
  and\ \citenamefont {Melchiorri}}]{re2}%
  \BibitemOpen
  \bibfield  {author} {\bibinfo {author} {\bibfnamefont {Weiqiang}\
  \bibnamefont {Yang}}, \bibinfo {author} {\bibfnamefont {Supriya}\
  \bibnamefont {Pan}}, \bibinfo {author} {\bibfnamefont {Eleonora}\
  \bibnamefont {Di~Valentino}}, \bibinfo {author} {\bibfnamefont {Olga}\
  \bibnamefont {Mena}}, \ and\ \bibinfo {author} {\bibfnamefont {Alessandro}\
  \bibnamefont {Melchiorri}},\ }\bibfield  {title} {\enquote {\bibinfo {title}
  {{2021-$H_0$ Odyssey: Closed, Phantom and Interacting Dark Energy
  Cosmologies}},}\ }\href@noop {} {\  (\bibinfo {year} {2021})},\ \Eprint
  {http://arxiv.org/abs/2101.03129} {arXiv:2101.03129 [astro-ph.CO]}
  \BibitemShut {NoStop}%
\bibitem [{\citenamefont {Di~Valentino}\ \emph
  {et~al.}(2020{\natexlab{a}})\citenamefont {Di~Valentino} \emph
  {et~al.}}]{re3}%
  \BibitemOpen
  \bibfield  {author} {\bibinfo {author} {\bibfnamefont {Eleonora}\
  \bibnamefont {Di~Valentino}} \emph {et~al.},\ }\bibfield  {title} {\enquote
  {\bibinfo {title} {{Cosmology Intertwined I: Perspectives for the Next
  Decade}},}\ }\href@noop {} {\  (\bibinfo {year} {2020}{\natexlab{a}})},\
  \Eprint {http://arxiv.org/abs/2008.11283} {arXiv:2008.11283 [astro-ph.CO]}
  \BibitemShut {NoStop}%
\bibitem [{\citenamefont {Di~Valentino}\ \emph
  {et~al.}(2020{\natexlab{b}})\citenamefont {Di~Valentino} \emph
  {et~al.}}]{re4}%
  \BibitemOpen
  \bibfield  {author} {\bibinfo {author} {\bibfnamefont {Eleonora}\
  \bibnamefont {Di~Valentino}} \emph {et~al.},\ }\bibfield  {title} {\enquote
  {\bibinfo {title} {{Cosmology Intertwined III: $f \sigma_8$ and $S_8$}},}\
  }\href@noop {} {\  (\bibinfo {year} {2020}{\natexlab{b}})},\ \Eprint
  {http://arxiv.org/abs/2008.11285} {arXiv:2008.11285 [astro-ph.CO]}
  \BibitemShut {NoStop}%
\bibitem [{\citenamefont {Yang}\ \emph {et~al.}(2020)\citenamefont {Yang},
  \citenamefont {Di~Valentino}, \citenamefont {Mena}, \citenamefont {Pan},\
  and\ \citenamefont {Nunes}}]{re5}%
  \BibitemOpen
  \bibfield  {author} {\bibinfo {author} {\bibfnamefont {Weiqiang}\
  \bibnamefont {Yang}}, \bibinfo {author} {\bibfnamefont {Eleonora}\
  \bibnamefont {Di~Valentino}}, \bibinfo {author} {\bibfnamefont {Olga}\
  \bibnamefont {Mena}}, \bibinfo {author} {\bibfnamefont {Supriya}\
  \bibnamefont {Pan}}, \ and\ \bibinfo {author} {\bibfnamefont {Rafael~C.}\
  \bibnamefont {Nunes}},\ }\bibfield  {title} {\enquote {\bibinfo {title}
  {{All-inclusive interacting dark sector cosmologies}},}\ }\href {\doibase
  10.1103/PhysRevD.101.083509} {\bibfield  {journal} {\bibinfo  {journal}
  {Phys. Rev. D}\ }\textbf {\bibinfo {volume} {101}},\ \bibinfo {pages}
  {083509} (\bibinfo {year} {2020})},\ \Eprint
  {http://arxiv.org/abs/2001.10852} {arXiv:2001.10852 [astro-ph.CO]}
  \BibitemShut {NoStop}%
\bibitem [{\citenamefont {Harko}\ \emph {et~al.}(2018)\citenamefont {Harko},
  \citenamefont {Koivisto}, \citenamefont {Lobo}, \citenamefont {Olmo},\ and\
  \citenamefont {Rubiera-Garcia}}]{Harko:2018gxr}%
  \BibitemOpen
  \bibfield  {author} {\bibinfo {author} {\bibfnamefont {Tiberiu}\ \bibnamefont
  {Harko}}, \bibinfo {author} {\bibfnamefont {Tomi~S.}\ \bibnamefont
  {Koivisto}}, \bibinfo {author} {\bibfnamefont {Francisco S.~N.}\ \bibnamefont
  {Lobo}}, \bibinfo {author} {\bibfnamefont {Gonzalo~J.}\ \bibnamefont {Olmo}},
  \ and\ \bibinfo {author} {\bibfnamefont {Diego}\ \bibnamefont
  {Rubiera-Garcia}},\ }\bibfield  {title} {\enquote {\bibinfo {title}
  {{Coupling matter in modified $Q$ gravity}},}\ }\href {\doibase
  10.1103/PhysRevD.98.084043} {\bibfield  {journal} {\bibinfo  {journal} {Phys.
  Rev. D}\ }\textbf {\bibinfo {volume} {98}},\ \bibinfo {pages} {084043}
  (\bibinfo {year} {2018})},\ \Eprint {http://arxiv.org/abs/1806.10437}
  {arXiv:1806.10437 [gr-qc]} \BibitemShut {NoStop}%
\bibitem [{\citenamefont {Lazkoz}\ \emph {et~al.}(2019)\citenamefont {Lazkoz},
  \citenamefont {Lobo}, \citenamefont {Ortiz-Ba\~nos},\ and\ \citenamefont
  {Salzano}}]{Lazkoz:2019sjl}%
  \BibitemOpen
  \bibfield  {author} {\bibinfo {author} {\bibfnamefont {Ruth}\ \bibnamefont
  {Lazkoz}}, \bibinfo {author} {\bibfnamefont {Francisco S.~N.}\ \bibnamefont
  {Lobo}}, \bibinfo {author} {\bibfnamefont {Mar\'\i{}a}\ \bibnamefont
  {Ortiz-Ba\~nos}}, \ and\ \bibinfo {author} {\bibfnamefont {Vincenzo}\
  \bibnamefont {Salzano}},\ }\bibfield  {title} {\enquote {\bibinfo {title}
  {{Observational constraints of $f(Q)$ gravity}},}\ }\href {\doibase
  10.1103/PhysRevD.100.104027} {\bibfield  {journal} {\bibinfo  {journal}
  {Phys. Rev. D}\ }\textbf {\bibinfo {volume} {100}},\ \bibinfo {pages}
  {104027} (\bibinfo {year} {2019})},\ \Eprint
  {http://arxiv.org/abs/1907.13219} {arXiv:1907.13219 [gr-qc]} \BibitemShut
  {NoStop}%
\bibitem [{\citenamefont {Ayuso}\ \emph {et~al.}(2021)\citenamefont {Ayuso},
  \citenamefont {Lazkoz},\ and\ \citenamefont {Salzano}}]{Ayuso:2020dcu}%
  \BibitemOpen
  \bibfield  {author} {\bibinfo {author} {\bibfnamefont {Ismael}\ \bibnamefont
  {Ayuso}}, \bibinfo {author} {\bibfnamefont {Ruth}\ \bibnamefont {Lazkoz}}, \
  and\ \bibinfo {author} {\bibfnamefont {Vincenzo}\ \bibnamefont {Salzano}},\
  }\bibfield  {title} {\enquote {\bibinfo {title} {{Observational constraints
  on cosmological solutions of $f(Q)$ theories}},}\ }\href {\doibase
  10.1103/PhysRevD.103.063505} {\bibfield  {journal} {\bibinfo  {journal}
  {Phys. Rev. D}\ }\textbf {\bibinfo {volume} {103}},\ \bibinfo {pages}
  {063505} (\bibinfo {year} {2021})},\ \Eprint
  {http://arxiv.org/abs/2012.00046} {arXiv:2012.00046 [astro-ph.CO]}
  \BibitemShut {NoStop}%
\bibitem [{\citenamefont {Mandal}\ \emph {et~al.}(2020)\citenamefont {Mandal},
  \citenamefont {Sahoo},\ and\ \citenamefont {Santos}}]{Mandal:2020lyq}%
  \BibitemOpen
  \bibfield  {author} {\bibinfo {author} {\bibfnamefont {Sanjay}\ \bibnamefont
  {Mandal}}, \bibinfo {author} {\bibfnamefont {P.~K.}\ \bibnamefont {Sahoo}}, \
  and\ \bibinfo {author} {\bibfnamefont {J.~R.~L.}\ \bibnamefont {Santos}},\
  }\bibfield  {title} {\enquote {\bibinfo {title} {{Energy conditions in $f(Q)$
  gravity}},}\ }\href {\doibase 10.1103/PhysRevD.102.024057} {\bibfield
  {journal} {\bibinfo  {journal} {Phys. Rev. D}\ }\textbf {\bibinfo {volume}
  {102}},\ \bibinfo {pages} {024057} (\bibinfo {year} {2020})},\ \Eprint
  {http://arxiv.org/abs/2008.01563} {arXiv:2008.01563 [gr-qc]} \BibitemShut
  {NoStop}%
\bibitem [{\citenamefont {J\"arv}\ \emph {et~al.}(2018)\citenamefont {J\"arv},
  \citenamefont {R\"unkla}, \citenamefont {Saal},\ and\ \citenamefont
  {Vilson}}]{Jarv:2018bgs}%
  \BibitemOpen
  \bibfield  {author} {\bibinfo {author} {\bibfnamefont {Laur}\ \bibnamefont
  {J\"arv}}, \bibinfo {author} {\bibfnamefont {Mihkel}\ \bibnamefont
  {R\"unkla}}, \bibinfo {author} {\bibfnamefont {Margus}\ \bibnamefont {Saal}},
  \ and\ \bibinfo {author} {\bibfnamefont {Ott}\ \bibnamefont {Vilson}},\
  }\bibfield  {title} {\enquote {\bibinfo {title} {{Nonmetricity formulation of
  general relativity and its scalar-tensor extension}},}\ }\href {\doibase
  10.1103/PhysRevD.97.124025} {\bibfield  {journal} {\bibinfo  {journal} {Phys.
  Rev. D}\ }\textbf {\bibinfo {volume} {97}},\ \bibinfo {pages} {124025}
  (\bibinfo {year} {2018})},\ \Eprint {http://arxiv.org/abs/1802.00492}
  {arXiv:1802.00492 [gr-qc]} \BibitemShut {NoStop}%
\bibitem [{\citenamefont {Golovnev}\ and\ \citenamefont
  {Koivisto}(2018)}]{Golovnev:2018wbh}%
  \BibitemOpen
  \bibfield  {author} {\bibinfo {author} {\bibfnamefont {Alexey}\ \bibnamefont
  {Golovnev}}\ and\ \bibinfo {author} {\bibfnamefont {Tomi}\ \bibnamefont
  {Koivisto}},\ }\bibfield  {title} {\enquote {\bibinfo {title} {{Cosmological
  perturbations in modified teleparallel gravity models}},}\ }\href {\doibase
  10.1088/1475-7516/2018/11/012} {\bibfield  {journal} {\bibinfo  {journal}
  {JCAP}\ }\textbf {\bibinfo {volume} {11}},\ \bibinfo {pages} {012} (\bibinfo
  {year} {2018})},\ \Eprint {http://arxiv.org/abs/1808.05565} {arXiv:1808.05565
  [gr-qc]} \BibitemShut {NoStop}%
\bibitem [{\citenamefont {Khyllep}\ and\ \citenamefont
  {Dutta}(2021)}]{Khyllep:2021yyp}%
  \BibitemOpen
  \bibfield  {author} {\bibinfo {author} {\bibfnamefont {Wompherdeiki}\
  \bibnamefont {Khyllep}}\ and\ \bibinfo {author} {\bibfnamefont {Jibitesh}\
  \bibnamefont {Dutta}},\ }\bibfield  {title} {\enquote {\bibinfo {title}
  {{Cosmological dynamics and bifurcation analysis of the general non-minimally
  coupled scalar field models}},}\ }\href@noop {} {\  (\bibinfo {year}
  {2021})},\ \Eprint {http://arxiv.org/abs/2102.04744} {arXiv:2102.04744
  [gr-qc]} \BibitemShut {NoStop}%
\bibitem [{\citenamefont {Dutta}\ \emph {et~al.}(2020)\citenamefont {Dutta},
  \citenamefont {Jarv}, \citenamefont {Khyllep},\ and\ \citenamefont
  {Tokke}}]{Dutta:2020uha}%
  \BibitemOpen
  \bibfield  {author} {\bibinfo {author} {\bibfnamefont {Jibitesh}\
  \bibnamefont {Dutta}}, \bibinfo {author} {\bibfnamefont {Laur}\ \bibnamefont
  {Jarv}}, \bibinfo {author} {\bibfnamefont {Wompherdeiki}\ \bibnamefont
  {Khyllep}}, \ and\ \bibinfo {author} {\bibfnamefont {Sulev}\ \bibnamefont
  {Tokke}},\ }\bibfield  {title} {\enquote {\bibinfo {title} {{From inflation
  to dark energy in scalar-tensor cosmology}},}\ }\href@noop {} {\  (\bibinfo
  {year} {2020})},\ \Eprint {http://arxiv.org/abs/2007.06601} {arXiv:2007.06601
  [gr-qc]} \BibitemShut {NoStop}%
\bibitem [{\citenamefont {Alho}\ \emph {et~al.}(2020)\citenamefont {Alho},
  \citenamefont {Uggla},\ and\ \citenamefont {Wainwright}}]{Alho:2020cdg}%
  \BibitemOpen
  \bibfield  {author} {\bibinfo {author} {\bibfnamefont {Artur}\ \bibnamefont
  {Alho}}, \bibinfo {author} {\bibfnamefont {Claes}\ \bibnamefont {Uggla}}, \
  and\ \bibinfo {author} {\bibfnamefont {John}\ \bibnamefont {Wainwright}},\
  }\bibfield  {title} {\enquote {\bibinfo {title} {{Dynamical systems in
  perturbative scalar field cosmology}},}\ }\href {\doibase
  10.1088/1361-6382/abb73a} {\bibfield  {journal} {\bibinfo  {journal} {Class.
  Quant. Grav.}\ }\textbf {\bibinfo {volume} {37}},\ \bibinfo {pages} {225011}
  (\bibinfo {year} {2020})},\ \Eprint {http://arxiv.org/abs/2006.00800}
  {arXiv:2006.00800 [gr-qc]} \BibitemShut {NoStop}%
\bibitem [{\citenamefont {Leon}\ \emph {et~al.}(2020)\citenamefont {Leon},
  \citenamefont {Paliathanasis},\ and\ \citenamefont {Dimakis}}]{Leon:2020cge}%
  \BibitemOpen
  \bibfield  {author} {\bibinfo {author} {\bibfnamefont {Genly}\ \bibnamefont
  {Leon}}, \bibinfo {author} {\bibfnamefont {Andronikos}\ \bibnamefont
  {Paliathanasis}}, \ and\ \bibinfo {author} {\bibfnamefont {N.}~\bibnamefont
  {Dimakis}},\ }\bibfield  {title} {\enquote {\bibinfo {title} {{Exact
  Kantowski\textendash{}Sachs spacetimes in Einstein\textendash{}Aether scalar
  field theory}},}\ }\href {\doibase 10.1140/epjc/s10052-020-08721-1}
  {\bibfield  {journal} {\bibinfo  {journal} {Eur. Phys. J. C}\ }\textbf
  {\bibinfo {volume} {80}},\ \bibinfo {pages} {1149} (\bibinfo {year}
  {2020})},\ \Eprint {http://arxiv.org/abs/2010.02775} {arXiv:2010.02775
  [gr-qc]} \BibitemShut {NoStop}%
\bibitem [{\citenamefont {Giacomini}\ \emph {et~al.}(2020)\citenamefont
  {Giacomini}, \citenamefont {Leon}, \citenamefont {Paliathanasis},\ and\
  \citenamefont {Pan}}]{Giacomini:2020zmv}%
  \BibitemOpen
  \bibfield  {author} {\bibinfo {author} {\bibfnamefont {Alex}\ \bibnamefont
  {Giacomini}}, \bibinfo {author} {\bibfnamefont {Genly}\ \bibnamefont {Leon}},
  \bibinfo {author} {\bibfnamefont {Andronikos}\ \bibnamefont {Paliathanasis}},
  \ and\ \bibinfo {author} {\bibfnamefont {Supriya}\ \bibnamefont {Pan}},\
  }\bibfield  {title} {\enquote {\bibinfo {title} {{Dynamics of Quintessence in
  Generalized Uncertainty Principle}},}\ }\href {\doibase
  10.1140/epjc/s10052-020-08508-4} {\bibfield  {journal} {\bibinfo  {journal}
  {Eur. Phys. J. C}\ }\textbf {\bibinfo {volume} {80}},\ \bibinfo {pages} {931}
  (\bibinfo {year} {2020})},\ \Eprint {http://arxiv.org/abs/2008.01395}
  {arXiv:2008.01395 [gr-qc]} \BibitemShut {NoStop}%
\bibitem [{\citenamefont {Paliathanasis}(2020)}]{Paliathanasis:2020wjl}%
  \BibitemOpen
  \bibfield  {author} {\bibinfo {author} {\bibfnamefont {Andronikos}\
  \bibnamefont {Paliathanasis}},\ }\bibfield  {title} {\enquote {\bibinfo
  {title} {{Dynamics of Chiral Cosmology}},}\ }\href {\doibase
  10.1088/1361-6382/aba667} {\bibfield  {journal} {\bibinfo  {journal} {Class.
  Quant. Grav.}\ }\textbf {\bibinfo {volume} {37}},\ \bibinfo {pages} {195014}
  (\bibinfo {year} {2020})},\ \Eprint {http://arxiv.org/abs/2003.05342}
  {arXiv:2003.05342 [gr-qc]} \BibitemShut {NoStop}%
\bibitem [{\citenamefont {Christodoulidis}\ \emph {et~al.}(2019)\citenamefont
  {Christodoulidis}, \citenamefont {Roest},\ and\ \citenamefont
  {Sfakianakis}}]{Christodoulidis:2019jsx}%
  \BibitemOpen
  \bibfield  {author} {\bibinfo {author} {\bibfnamefont {Perseas}\ \bibnamefont
  {Christodoulidis}}, \bibinfo {author} {\bibfnamefont {Diederik}\ \bibnamefont
  {Roest}}, \ and\ \bibinfo {author} {\bibfnamefont {Evangelos~I.}\
  \bibnamefont {Sfakianakis}},\ }\bibfield  {title} {\enquote {\bibinfo {title}
  {{Scaling attractors in multi-field inflation}},}\ }\href {\doibase
  10.1088/1475-7516/2019/12/059} {\bibfield  {journal} {\bibinfo  {journal}
  {JCAP}\ }\textbf {\bibinfo {volume} {12}},\ \bibinfo {pages} {059} (\bibinfo
  {year} {2019})},\ \Eprint {http://arxiv.org/abs/1903.06116} {arXiv:1903.06116
  [hep-th]} \BibitemShut {NoStop}%
\bibitem [{\citenamefont {Coley}\ and\ \citenamefont
  {Leon}(2019)}]{Coley:2019tyx}%
  \BibitemOpen
  \bibfield  {author} {\bibinfo {author} {\bibfnamefont {Alan}\ \bibnamefont
  {Coley}}\ and\ \bibinfo {author} {\bibfnamefont {Genly}\ \bibnamefont
  {Leon}},\ }\bibfield  {title} {\enquote {\bibinfo {title} {{Static
  Spherically Symmetric Einstein-aether models I: Perfect fluids with a linear
  equation of state and scalar fields with an exponential self-interacting
  potential}},}\ }\href {\doibase 10.1007/s10714-019-2598-y} {\bibfield
  {journal} {\bibinfo  {journal} {Gen. Rel. Grav.}\ }\textbf {\bibinfo {volume}
  {51}},\ \bibinfo {pages} {115} (\bibinfo {year} {2019})},\ \Eprint
  {http://arxiv.org/abs/1905.02003} {arXiv:1905.02003 [gr-qc]} \BibitemShut
  {NoStop}%
\bibitem [{\citenamefont {Basilakos}\ \emph {et~al.}(2019)\citenamefont
  {Basilakos}, \citenamefont {Leon}, \citenamefont {Papagiannopoulos},\ and\
  \citenamefont {Saridakis}}]{Basilakos:2019dof}%
  \BibitemOpen
  \bibfield  {author} {\bibinfo {author} {\bibfnamefont {Spyros}\ \bibnamefont
  {Basilakos}}, \bibinfo {author} {\bibfnamefont {Genly}\ \bibnamefont {Leon}},
  \bibinfo {author} {\bibfnamefont {G.}~\bibnamefont {Papagiannopoulos}}, \
  and\ \bibinfo {author} {\bibfnamefont {Emmanuel~N.}\ \bibnamefont
  {Saridakis}},\ }\bibfield  {title} {\enquote {\bibinfo {title} {{Dynamical
  system analysis at background and perturbation levels: Quintessence in severe
  disadvantage comparing to $\Lambda$CDM}},}\ }\href {\doibase
  10.1103/PhysRevD.100.043524} {\bibfield  {journal} {\bibinfo  {journal}
  {Phys. Rev. D}\ }\textbf {\bibinfo {volume} {100}},\ \bibinfo {pages}
  {043524} (\bibinfo {year} {2019})},\ \Eprint
  {http://arxiv.org/abs/1904.01563} {arXiv:1904.01563 [gr-qc]} \BibitemShut
  {NoStop}%
\bibitem [{\citenamefont {Cid}\ \emph {et~al.}(2018)\citenamefont {Cid},
  \citenamefont {Izaurieta}, \citenamefont {Leon}, \citenamefont {Medina},\
  and\ \citenamefont {Narbona}}]{Cid:2017wtf}%
  \BibitemOpen
  \bibfield  {author} {\bibinfo {author} {\bibfnamefont {Antonella}\
  \bibnamefont {Cid}}, \bibinfo {author} {\bibfnamefont {Fernando}\
  \bibnamefont {Izaurieta}}, \bibinfo {author} {\bibfnamefont {Genly}\
  \bibnamefont {Leon}}, \bibinfo {author} {\bibfnamefont {Perla}\ \bibnamefont
  {Medina}}, \ and\ \bibinfo {author} {\bibfnamefont {Daniela}\ \bibnamefont
  {Narbona}},\ }\bibfield  {title} {\enquote {\bibinfo {title} {{Non-minimally
  coupled scalar field cosmology with torsion}},}\ }\href {\doibase
  10.1088/1475-7516/2018/04/041} {\bibfield  {journal} {\bibinfo  {journal}
  {JCAP}\ }\textbf {\bibinfo {volume} {04}},\ \bibinfo {pages} {041} (\bibinfo
  {year} {2018})},\ \Eprint {http://arxiv.org/abs/1704.04563} {arXiv:1704.04563
  [gr-qc]} \BibitemShut {NoStop}%
\bibitem [{\citenamefont {Bahamonde}\ \emph {et~al.}(2018)\citenamefont
  {Bahamonde}, \citenamefont {B\"ohmer}, \citenamefont {Carloni}, \citenamefont
  {Copeland}, \citenamefont {Fang},\ and\ \citenamefont
  {Tamanini}}]{Bahamonde:2017ize}%
  \BibitemOpen
  \bibfield  {author} {\bibinfo {author} {\bibfnamefont {Sebastian}\
  \bibnamefont {Bahamonde}}, \bibinfo {author} {\bibfnamefont {Christian~G.}\
  \bibnamefont {B\"ohmer}}, \bibinfo {author} {\bibfnamefont {Sante}\
  \bibnamefont {Carloni}}, \bibinfo {author} {\bibfnamefont {Edmund~J.}\
  \bibnamefont {Copeland}}, \bibinfo {author} {\bibfnamefont {Wei}\
  \bibnamefont {Fang}}, \ and\ \bibinfo {author} {\bibfnamefont {Nicola}\
  \bibnamefont {Tamanini}},\ }\bibfield  {title} {\enquote {\bibinfo {title}
  {{Dynamical systems applied to cosmology: dark energy and modified
  gravity}},}\ }\href {\doibase 10.1016/j.physrep.2018.09.001} {\bibfield
  {journal} {\bibinfo  {journal} {Phys. Rept.}\ }\textbf {\bibinfo {volume}
  {775-777}},\ \bibinfo {pages} {1--122} (\bibinfo {year} {2018})},\ \Eprint
  {http://arxiv.org/abs/1712.03107} {arXiv:1712.03107 [gr-qc]} \BibitemShut
  {NoStop}%
\bibitem [{\citenamefont {Carloni}(2015)}]{Carloni:2015jla}%
  \BibitemOpen
  \bibfield  {author} {\bibinfo {author} {\bibfnamefont {Sante}\ \bibnamefont
  {Carloni}},\ }\bibfield  {title} {\enquote {\bibinfo {title} {{A new approach
  to the analysis of the phase space of $f(R)$-gravity}},}\ }\href {\doibase
  10.1088/1475-7516/2015/09/013} {\bibfield  {journal} {\bibinfo  {journal}
  {JCAP}\ }\textbf {\bibinfo {volume} {09}},\ \bibinfo {pages} {013} (\bibinfo
  {year} {2015})},\ \Eprint {http://arxiv.org/abs/1505.06015} {arXiv:1505.06015
  [gr-qc]} \BibitemShut {NoStop}%
\bibitem [{\citenamefont {Alho}\ \emph {et~al.}(2016)\citenamefont {Alho},
  \citenamefont {Carloni},\ and\ \citenamefont {Uggla}}]{Alho:2016gzi}%
  \BibitemOpen
  \bibfield  {author} {\bibinfo {author} {\bibfnamefont {Artur}\ \bibnamefont
  {Alho}}, \bibinfo {author} {\bibfnamefont {Sante}\ \bibnamefont {Carloni}}, \
  and\ \bibinfo {author} {\bibfnamefont {Claes}\ \bibnamefont {Uggla}},\
  }\bibfield  {title} {\enquote {\bibinfo {title} {{On dynamical systems
  approaches and methods in $f(R)$ cosmology}},}\ }\href {\doibase
  10.1088/1475-7516/2016/08/064} {\bibfield  {journal} {\bibinfo  {journal}
  {JCAP}\ }\textbf {\bibinfo {volume} {08}},\ \bibinfo {pages} {064} (\bibinfo
  {year} {2016})},\ \Eprint {http://arxiv.org/abs/1607.05715} {arXiv:1607.05715
  [gr-qc]} \BibitemShut {NoStop}%
\bibitem [{\citenamefont {Hohmann}\ \emph {et~al.}(2017)\citenamefont
  {Hohmann}, \citenamefont {Jarv},\ and\ \citenamefont
  {Ualikhanova}}]{Hohmann:2017jao}%
  \BibitemOpen
  \bibfield  {author} {\bibinfo {author} {\bibfnamefont {Manuel}\ \bibnamefont
  {Hohmann}}, \bibinfo {author} {\bibfnamefont {Laur}\ \bibnamefont {Jarv}}, \
  and\ \bibinfo {author} {\bibfnamefont {Ulbossyn}\ \bibnamefont
  {Ualikhanova}},\ }\bibfield  {title} {\enquote {\bibinfo {title} {{Dynamical
  systems approach and generic properties of $f(T)$ cosmology}},}\ }\href
  {\doibase 10.1103/PhysRevD.96.043508} {\bibfield  {journal} {\bibinfo
  {journal} {Phys. Rev. D}\ }\textbf {\bibinfo {volume} {96}},\ \bibinfo
  {pages} {043508} (\bibinfo {year} {2017})},\ \Eprint
  {http://arxiv.org/abs/1706.02376} {arXiv:1706.02376 [gr-qc]} \BibitemShut
  {NoStop}%
\bibitem [{\citenamefont {Tsamparlis}\ and\ \citenamefont
  {Paliathanasis}(2018)}]{mtsa1}%
  \BibitemOpen
  \bibfield  {author} {\bibinfo {author} {\bibfnamefont {Michael}\ \bibnamefont
  {Tsamparlis}}\ and\ \bibinfo {author} {\bibfnamefont {Andronikos}\
  \bibnamefont {Paliathanasis}},\ }\bibfield  {title} {\enquote {\bibinfo
  {title} {{Symmetries of Differential Equations in Cosmology}},}\ }\href
  {\doibase 10.3390/sym10070233} {\bibfield  {journal} {\bibinfo  {journal}
  {Symmetry}\ }\textbf {\bibinfo {volume} {10}},\ \bibinfo {pages} {233}
  (\bibinfo {year} {2018})},\ \Eprint {http://arxiv.org/abs/1806.05888}
  {arXiv:1806.05888 [gr-qc]} \BibitemShut {NoStop}%
\bibitem [{\citenamefont {Kovalevskaya}(1889)}]{sof}%
  \BibitemOpen
  \bibfield  {author} {\bibinfo {author} {\bibfnamefont {S.}~\bibnamefont
  {Kovalevskaya}},\ }\bibfield  {title} {\enquote {\bibinfo {title} {{Sur le
  probl\'eme de la rotation d'un corps solide autour d'un point fixe}},}\
  }\href@noop {} {\bibfield  {journal} {\bibinfo  {journal} {Acta Mathematica}\
  }\textbf {\bibinfo {volume} {12}},\ \bibinfo {pages} {177} (\bibinfo {year}
  {1889})}\BibitemShut {NoStop}%
\bibitem [{\citenamefont {Painlev\'e}()}]{pain1}%
  \BibitemOpen
  \bibfield  {author} {\bibinfo {author} {\bibfnamefont {P.}~\bibnamefont
  {Painlev\'e}},\ }\bibfield  {title} {\enquote {\bibinfo {title} {{Le\c cons
  sur la th\'eorie analytique des \'equations diff\'erentielles}},}\
  }\href@noop {} {\bibinfo  {journal} {(Le\c cons de Stockholm, 1895) (Hermann,
  Paris, 1897). Reprinted, \oe uvres de Paul Painlev\'e, vol.~I, \'Editions du
  CNRS, Paris, 1973.}\ }\BibitemShut {NoStop}%
\bibitem [{\citenamefont {Painlev\'e}(1900)}]{pain2}%
  \BibitemOpen
\bibfield  {journal} {  }\bibfield  {author} {\bibinfo {author} {\bibfnamefont
  {Paul}\ \bibnamefont {Painlev\'e}},\ }\bibfield  {title} {\enquote {\bibinfo
  {title} {M\'emoire sur les \'equations diff\'erentielles dont l'int\'egrale
  g\'en\'erale est uniforme},}\ }\href {\doibase 10.24033/bsmf.633} {\bibfield
  {journal} {\bibinfo  {journal} {Bulletin de la Soci\'et\'e Math\'ematique de
  France}\ }\textbf {\bibinfo {volume} {28}},\ \bibinfo {pages} {201--261}
  (\bibinfo {year} {1900})}\BibitemShut {NoStop}%
\bibitem [{\citenamefont {Painlevé}(1902)}]{pain3}%
  \BibitemOpen
  \bibfield  {author} {\bibinfo {author} {\bibfnamefont {P.}~\bibnamefont
  {Painlevé}},\ }\bibfield  {title} {\enquote {\bibinfo {title} {{Sur les
  équations différentielles du second ordre et d'ordre supérieur dont
  l'intégrale générale est uniforme}},}\ }\href {\doibase
  10.1007/BF02419020} {\bibfield  {journal} {\bibinfo  {journal} {Acta
  Mathematica}\ }\textbf {\bibinfo {volume} {25}},\ \bibinfo {pages} {1 -- 85}
  (\bibinfo {year} {1902})}\BibitemShut {NoStop}%
\bibitem [{\citenamefont {Painlev\'e}(1906)}]{pain4}%
  \BibitemOpen
  \bibfield  {author} {\bibinfo {author} {\bibfnamefont {P.}~\bibnamefont
  {Painlev\'e}},\ }\bibfield  {title} {\enquote {\bibinfo {title} {{Sur les
  \'equations diff\'erentielles du second ordre \'a points critiques fixes}},}\
  }\href@noop {} {\bibfield  {journal} {\bibinfo  {journal} {Comptes Rendus de
  la Acad\'emie des Sciences de Paris}\ }\textbf {\bibinfo {volume} {143}},\
  \bibinfo {pages} {1111--1117} (\bibinfo {year} {1906})}\BibitemShut {NoStop}%
\bibitem [{\citenamefont {Cotsakis}\ and\ \citenamefont
  {Leach}(1994)}]{Cotsakis_1994}%
  \BibitemOpen
  \bibfield  {author} {\bibinfo {author} {\bibfnamefont {S.}~\bibnamefont
  {Cotsakis}}\ and\ \bibinfo {author} {\bibfnamefont {P.G.L}\ \bibnamefont
  {Leach}},\ }\bibfield  {title} {\enquote {\bibinfo {title} {Painleve analysis
  of the mixmaster universe},}\ }\href {\doibase 10.1088/0305-4470/27/5/026}
  {\bibfield  {journal} {\bibinfo  {journal} {Journal of Physics A:
  Mathematical and General}\ }\textbf {\bibinfo {volume} {27}},\ \bibinfo
  {pages} {1625--1631} (\bibinfo {year} {1994})}\BibitemShut {NoStop}%
\bibitem [{\citenamefont {Christiansen}\ \emph {et~al.}(1995)\citenamefont
  {Christiansen}, \citenamefont {Rugh},\ and\ \citenamefont
  {Rugh}}]{Christiansen_1995}%
  \BibitemOpen
  \bibfield  {author} {\bibinfo {author} {\bibfnamefont {F}~\bibnamefont
  {Christiansen}}, \bibinfo {author} {\bibfnamefont {H~H}\ \bibnamefont
  {Rugh}}, \ and\ \bibinfo {author} {\bibfnamefont {S~E}\ \bibnamefont
  {Rugh}},\ }\bibfield  {title} {\enquote {\bibinfo {title} {Non-integrability
  of the mixmaster universe},}\ }\href {\doibase 10.1088/0305-4470/28/3/019}
  {\bibfield  {journal} {\bibinfo  {journal} {Journal of Physics A:
  Mathematical and General}\ }\textbf {\bibinfo {volume} {28}},\ \bibinfo
  {pages} {657--667} (\bibinfo {year} {1995})}\BibitemShut {NoStop}%
\bibitem [{\citenamefont {Demaret}\ and\ \citenamefont
  {Scheen}(1996)}]{Demaret_1996}%
  \BibitemOpen
  \bibfield  {author} {\bibinfo {author} {\bibfnamefont {Jacques}\ \bibnamefont
  {Demaret}}\ and\ \bibinfo {author} {\bibfnamefont {Christian}\ \bibnamefont
  {Scheen}},\ }\bibfield  {title} {\enquote {\bibinfo {title} {Painlev{\'{e}}
  singularity analysis of the perfect fluid bianchi type-{IX} relativistic
  cosmological model},}\ }\href {\doibase 10.1088/0305-4470/29/1/009}
  {\bibfield  {journal} {\bibinfo  {journal} {Journal of Physics A:
  Mathematical and General}\ }\textbf {\bibinfo {volume} {29}},\ \bibinfo
  {pages} {59--76} (\bibinfo {year} {1996})}\BibitemShut {NoStop}%
\bibitem [{\citenamefont {Helmi}\ and\ \citenamefont
  {Vucetich}(1997)}]{Helmi:1997mj}%
  \BibitemOpen
  \bibfield  {author} {\bibinfo {author} {\bibfnamefont {Amina}\ \bibnamefont
  {Helmi}}\ and\ \bibinfo {author} {\bibfnamefont {Hector}\ \bibnamefont
  {Vucetich}},\ }\bibfield  {title} {\enquote {\bibinfo {title}
  {{Nonintegrability and chaos in classical cosmology}},}\ }\href {\doibase
  10.1016/S0375-9601(97)00258-2} {\bibfield  {journal} {\bibinfo  {journal}
  {Phys. Lett. A}\ }\textbf {\bibinfo {volume} {230}},\ \bibinfo {pages} {153}
  (\bibinfo {year} {1997})},\ \Eprint {http://arxiv.org/abs/gr-qc/9705009}
  {arXiv:gr-qc/9705009} \BibitemShut {NoStop}%
\bibitem [{\citenamefont {Miritzis}\ \emph {et~al.}(2000)\citenamefont
  {Miritzis}, \citenamefont {Leach},\ and\ \citenamefont
  {Cotsakis}}]{Miritzis:2000js}%
  \BibitemOpen
  \bibfield  {author} {\bibinfo {author} {\bibfnamefont {John}\ \bibnamefont
  {Miritzis}}, \bibinfo {author} {\bibfnamefont {Peter}\ \bibnamefont {Leach}},
  \ and\ \bibinfo {author} {\bibfnamefont {Spiros}\ \bibnamefont {Cotsakis}},\
  }\bibfield  {title} {\enquote {\bibinfo {title} {{Symmetry, singularities and
  integrability in complex dynamics. 4. Painleve integrability of isotropic
  cosmologies}},}\ }\href@noop {} {\bibfield  {journal} {\bibinfo  {journal}
  {Grav. Cosmol.}\ }\textbf {\bibinfo {volume} {6}},\ \bibinfo {pages}
  {282--290} (\bibinfo {year} {2000})},\ \Eprint
  {http://arxiv.org/abs/gr-qc/0011019} {arXiv:gr-qc/0011019} \BibitemShut
  {NoStop}%
\bibitem [{\citenamefont {Leach}\ \emph {et~al.}(2001)\citenamefont {Leach},
  \citenamefont {Cotsakis},\ and\ \citenamefont {Miritzis}}]{Leach:2001aw}%
  \BibitemOpen
  \bibfield  {author} {\bibinfo {author} {\bibfnamefont {Peter}\ \bibnamefont
  {Leach}}, \bibinfo {author} {\bibfnamefont {Spiros}\ \bibnamefont
  {Cotsakis}}, \ and\ \bibinfo {author} {\bibfnamefont {John}\ \bibnamefont
  {Miritzis}},\ }\bibfield  {title} {\enquote {\bibinfo {title} {{Symmetry,
  singularities and integrability in complex dynamics. VI: Integrability
  properties of FRW scalar cosmologies}},}\ }\href@noop {} {\bibfield
  {journal} {\bibinfo  {journal} {Grav. Cosmol.}\ }\textbf {\bibinfo {volume}
  {7}},\ \bibinfo {pages} {311--320} (\bibinfo {year} {2001})},\ \Eprint
  {http://arxiv.org/abs/gr-qc/0107038} {arXiv:gr-qc/0107038} \BibitemShut
  {NoStop}%
\bibitem [{\citenamefont {Leon}\ \emph {et~al.}(2018)\citenamefont {Leon},
  \citenamefont {Paliathanasis},\ and\ \citenamefont
  {Morales-Mart\'\i{}nez}}]{Leon:2018lnd}%
  \BibitemOpen
  \bibfield  {author} {\bibinfo {author} {\bibfnamefont {Genly}\ \bibnamefont
  {Leon}}, \bibinfo {author} {\bibfnamefont {Andronikos}\ \bibnamefont
  {Paliathanasis}}, \ and\ \bibinfo {author} {\bibfnamefont {Jorge~Luis}\
  \bibnamefont {Morales-Mart\'\i{}nez}},\ }\bibfield  {title} {\enquote
  {\bibinfo {title} {{The past and future dynamics of quintom dark energy
  models}},}\ }\href {\doibase 10.1140/epjc/s10052-018-6225-y} {\bibfield
  {journal} {\bibinfo  {journal} {Eur. Phys. J. C}\ }\textbf {\bibinfo {volume}
  {78}},\ \bibinfo {pages} {753} (\bibinfo {year} {2018})},\ \Eprint
  {http://arxiv.org/abs/1808.05634} {arXiv:1808.05634 [gr-qc]} \BibitemShut
  {NoStop}%
\bibitem [{\citenamefont {Basilakos}\ \emph {et~al.}(2018)\citenamefont
  {Basilakos}, \citenamefont {Paliathanasis}, \citenamefont {Barrow},\ and\
  \citenamefont {Papagiannopoulos}}]{Basilakos:2018xjp}%
  \BibitemOpen
  \bibfield  {author} {\bibinfo {author} {\bibfnamefont {S.}~\bibnamefont
  {Basilakos}}, \bibinfo {author} {\bibfnamefont {A.}~\bibnamefont
  {Paliathanasis}}, \bibinfo {author} {\bibfnamefont {J.~D.}\ \bibnamefont
  {Barrow}}, \ and\ \bibinfo {author} {\bibfnamefont {G.}~\bibnamefont
  {Papagiannopoulos}},\ }\bibfield  {title} {\enquote {\bibinfo {title}
  {{Cosmological singularities and analytical solutions in varying vacuum
  cosmologies}},}\ }\href {\doibase 10.1140/epjc/s10052-018-6139-8} {\bibfield
  {journal} {\bibinfo  {journal} {Eur. Phys. J. C}\ }\textbf {\bibinfo {volume}
  {78}},\ \bibinfo {pages} {684} (\bibinfo {year} {2018})},\ \Eprint
  {http://arxiv.org/abs/1804.03656} {arXiv:1804.03656 [gr-qc]} \BibitemShut
  {NoStop}%
\bibitem [{\citenamefont {Paliathanasis}\ and\ \citenamefont
  {Leach}(2016{\natexlab{a}})}]{Paliathanasis:2016tch}%
  \BibitemOpen
  \bibfield  {author} {\bibinfo {author} {\bibfnamefont {Andronikos}\
  \bibnamefont {Paliathanasis}}\ and\ \bibinfo {author} {\bibfnamefont
  {P.~G.~L.}\ \bibnamefont {Leach}},\ }\bibfield  {title} {\enquote {\bibinfo
  {title} {{Analytical solutions in $R+qR^{n}$ cosmology from singularity
  analysis}},}\ }\href {\doibase 10.1016/j.physleta.2016.06.053} {\bibfield
  {journal} {\bibinfo  {journal} {Phys. Lett. A}\ }\textbf {\bibinfo {volume}
  {380}},\ \bibinfo {pages} {2815--2818} (\bibinfo {year}
  {2016}{\natexlab{a}})},\ \Eprint {http://arxiv.org/abs/1605.04204}
  {arXiv:1605.04204 [gr-qc]} \BibitemShut {NoStop}%
\bibitem [{\citenamefont {Paliathanasis}\ \emph {et~al.}(2016)\citenamefont
  {Paliathanasis}, \citenamefont {Barrow},\ and\ \citenamefont
  {Leach}}]{Paliathanasis:2016vsw}%
  \BibitemOpen
  \bibfield  {author} {\bibinfo {author} {\bibfnamefont {Andronikos}\
  \bibnamefont {Paliathanasis}}, \bibinfo {author} {\bibfnamefont {John~D.}\
  \bibnamefont {Barrow}}, \ and\ \bibinfo {author} {\bibfnamefont {P.~G.~L.}\
  \bibnamefont {Leach}},\ }\bibfield  {title} {\enquote {\bibinfo {title}
  {{Cosmological Solutions of $f(T)$ Gravity}},}\ }\href {\doibase
  10.1103/PhysRevD.94.023525} {\bibfield  {journal} {\bibinfo  {journal} {Phys.
  Rev. D}\ }\textbf {\bibinfo {volume} {94}},\ \bibinfo {pages} {023525}
  (\bibinfo {year} {2016})},\ \Eprint {http://arxiv.org/abs/1606.00659}
  {arXiv:1606.00659 [gr-qc]} \BibitemShut {NoStop}%
\bibitem [{\citenamefont {Cotsakis}\ \emph {et~al.}(2016)\citenamefont
  {Cotsakis}, \citenamefont {Kadry}, \citenamefont {Kolionis},\ and\
  \citenamefont {Tsokaros}}]{Cotsakis:2013aqa}%
  \BibitemOpen
  \bibfield  {author} {\bibinfo {author} {\bibfnamefont {Spiros}\ \bibnamefont
  {Cotsakis}}, \bibinfo {author} {\bibfnamefont {Seifedine}\ \bibnamefont
  {Kadry}}, \bibinfo {author} {\bibfnamefont {Georgios}\ \bibnamefont
  {Kolionis}}, \ and\ \bibinfo {author} {\bibfnamefont {Antonios}\ \bibnamefont
  {Tsokaros}},\ }\bibfield  {title} {\enquote {\bibinfo {title} {{Asymptotic
  vacua with higher derivatives}},}\ }\href {\doibase
  10.1016/j.physletb.2016.02.036} {\bibfield  {journal} {\bibinfo  {journal}
  {Phys. Lett. B}\ }\textbf {\bibinfo {volume} {755}},\ \bibinfo {pages}
  {387--392} (\bibinfo {year} {2016})},\ \Eprint
  {http://arxiv.org/abs/1303.2234} {arXiv:1303.2234 [gr-qc]} \BibitemShut
  {NoStop}%
\bibitem [{\citenamefont {Paliathanasis}(2017)}]{Paliathanasis:2017apr}%
  \BibitemOpen
  \bibfield  {author} {\bibinfo {author} {\bibfnamefont {Andronikos}\
  \bibnamefont {Paliathanasis}},\ }\bibfield  {title} {\enquote {\bibinfo
  {title} {{Analytic Solution of the Starobinsky Model for Inflation}},}\
  }\href {\doibase 10.1140/epjc/s10052-017-5009-0} {\bibfield  {journal}
  {\bibinfo  {journal} {Eur. Phys. J. C}\ }\textbf {\bibinfo {volume} {77}},\
  \bibinfo {pages} {438} (\bibinfo {year} {2017})},\ \Eprint
  {http://arxiv.org/abs/1706.06400} {arXiv:1706.06400 [gr-qc]} \BibitemShut
  {NoStop}%
\bibitem [{\citenamefont {Paliathanasis}\ and\ \citenamefont
  {Leon}(2020)}]{Paliathanasis:2019qch}%
  \BibitemOpen
  \bibfield  {author} {\bibinfo {author} {\bibfnamefont {Andronikos}\
  \bibnamefont {Paliathanasis}}\ and\ \bibinfo {author} {\bibfnamefont {Genly}\
  \bibnamefont {Leon}},\ }\bibfield  {title} {\enquote {\bibinfo {title}
  {Cosmological solutions in horava-lifshitz scalar field theory},}\ }\href
  {\doibase doi:10.1515/zna-2020-0003} {\bibfield  {journal} {\bibinfo
  {journal} {Zeitschrift f\"ur Naturforschung A}\ }\textbf {\bibinfo {volume}
  {75}},\ \bibinfo {pages} {523--532} (\bibinfo {year} {2020})}\BibitemShut
  {NoStop}%
\bibitem [{\citenamefont {Paliathanasis}\ and\ \citenamefont
  {Leach}(2017)}]{asze}%
  \BibitemOpen
  \bibfield  {author} {\bibinfo {author} {\bibfnamefont {Andronikos}\
  \bibnamefont {Paliathanasis}}\ and\ \bibinfo {author} {\bibfnamefont
  {P.~G.~L.}\ \bibnamefont {Leach}},\ }\bibfield  {title} {\enquote {\bibinfo
  {title} {{Symmetries and Singularities of the Szekeres System}},}\ }\href
  {\doibase 10.1016/j.physleta.2017.02.009} {\bibfield  {journal} {\bibinfo
  {journal} {Phys. Lett. A}\ }\textbf {\bibinfo {volume} {381}},\ \bibinfo
  {pages} {1277--1280} (\bibinfo {year} {2017})},\ \Eprint
  {http://arxiv.org/abs/1702.01593} {arXiv:1702.01593 [gr-qc]} \BibitemShut
  {NoStop}%
\bibitem [{\citenamefont {Dutta}\ \emph {et~al.}(2018)\citenamefont {Dutta},
  \citenamefont {Khyllep},\ and\ \citenamefont {Tamanini}}]{Dutta:2017wfd}%
  \BibitemOpen
  \bibfield  {author} {\bibinfo {author} {\bibfnamefont {Jibitesh}\
  \bibnamefont {Dutta}}, \bibinfo {author} {\bibfnamefont {Wompherdeiki}\
  \bibnamefont {Khyllep}}, \ and\ \bibinfo {author} {\bibfnamefont {Nicola}\
  \bibnamefont {Tamanini}},\ }\bibfield  {title} {\enquote {\bibinfo {title}
  {{Dark energy with a gradient coupling to the dark matter fluid: cosmological
  dynamics and structure formation}},}\ }\href {\doibase
  10.1088/1475-7516/2018/01/038} {\bibfield  {journal} {\bibinfo  {journal}
  {JCAP}\ }\textbf {\bibinfo {volume} {01}},\ \bibinfo {pages} {038} (\bibinfo
  {year} {2018})},\ \Eprint {http://arxiv.org/abs/1707.09246} {arXiv:1707.09246
  [gr-qc]} \BibitemShut {NoStop}%
\bibitem [{\citenamefont {Ishak}(2019)}]{Ishak:2018his}%
  \BibitemOpen
  \bibfield  {author} {\bibinfo {author} {\bibfnamefont {Mustapha}\
  \bibnamefont {Ishak}},\ }\bibfield  {title} {\enquote {\bibinfo {title}
  {{Testing General Relativity in Cosmology}},}\ }\href {\doibase
  10.1007/s41114-018-0017-4} {\bibfield  {journal} {\bibinfo  {journal} {Living
  Rev. Rel.}\ }\textbf {\bibinfo {volume} {22}},\ \bibinfo {pages} {1}
  (\bibinfo {year} {2019})},\ \Eprint {http://arxiv.org/abs/1806.10122}
  {arXiv:1806.10122 [astro-ph.CO]} \BibitemShut {NoStop}%
\bibitem [{\citenamefont {Basilakos}\ and\ \citenamefont
  {Anagnostopoulos}(2020)}]{bb1}%
  \BibitemOpen
  \bibfield  {author} {\bibinfo {author} {\bibfnamefont {Spyros}\ \bibnamefont
  {Basilakos}}\ and\ \bibinfo {author} {\bibfnamefont {Fotios~K.}\ \bibnamefont
  {Anagnostopoulos}},\ }\bibfield  {title} {\enquote {\bibinfo {title} {{Growth
  index of matter perturbations in the light of Dark Energy Survey}},}\ }\href
  {\doibase 10.1140/epjc/s10052-020-7770-8} {\bibfield  {journal} {\bibinfo
  {journal} {Eur. Phys. J. C}\ }\textbf {\bibinfo {volume} {80}},\ \bibinfo
  {pages} {212} (\bibinfo {year} {2020})},\ \Eprint
  {http://arxiv.org/abs/1903.10758} {arXiv:1903.10758 [astro-ph.CO]}
  \BibitemShut {NoStop}%
\bibitem [{\citenamefont {Khyllep}\ and\ \citenamefont
  {Dutta}(2019)}]{Khyllep:2019odd}%
  \BibitemOpen
  \bibfield  {author} {\bibinfo {author} {\bibfnamefont {Wompherdeiki}\
  \bibnamefont {Khyllep}}\ and\ \bibinfo {author} {\bibfnamefont {Jibitesh}\
  \bibnamefont {Dutta}},\ }\bibfield  {title} {\enquote {\bibinfo {title}
  {{Linear growth index of matter perturbations in Rastall gravity}},}\ }\href
  {\doibase 10.1016/j.physletb.2019.134796} {\bibfield  {journal} {\bibinfo
  {journal} {Phys. Lett. B}\ }\textbf {\bibinfo {volume} {797}},\ \bibinfo
  {pages} {134796} (\bibinfo {year} {2019})},\ \Eprint
  {http://arxiv.org/abs/1907.09221} {arXiv:1907.09221 [gr-qc]} \BibitemShut
  {NoStop}%
\bibitem [{\citenamefont {{Peebles}}(1993)}]{peeblesbook}%
  \BibitemOpen
  \bibfield  {author} {\bibinfo {author} {\bibfnamefont {P.~J.~E.}\
  \bibnamefont {{Peebles}}},\ }\href@noop {} {\emph {\bibinfo {title}
  {{Principles of Physical Cosmology}}}}\ (\bibinfo  {publisher} {Princeton
  University Press, Princeton},\ \bibinfo {year} {1993})\BibitemShut {NoStop}%
\bibitem [{\citenamefont {Barros}\ \emph {et~al.}(2020)\citenamefont {Barros},
  \citenamefont {Barreiro}, \citenamefont {Koivisto},\ and\ \citenamefont
  {Nunes}}]{Barros:2020bgg}%
  \BibitemOpen
  \bibfield  {author} {\bibinfo {author} {\bibfnamefont {Bruno~J.}\
  \bibnamefont {Barros}}, \bibinfo {author} {\bibfnamefont {Tiago}\
  \bibnamefont {Barreiro}}, \bibinfo {author} {\bibfnamefont {Tomi}\
  \bibnamefont {Koivisto}}, \ and\ \bibinfo {author} {\bibfnamefont
  {Nelson~J.}\ \bibnamefont {Nunes}},\ }\bibfield  {title} {\enquote {\bibinfo
  {title} {{Testing $F(Q)$ gravity with redshift space distortions}},}\ }\href
  {\doibase 10.1016/j.dark.2020.100616} {\bibfield  {journal} {\bibinfo
  {journal} {Phys. Dark Univ.}\ }\textbf {\bibinfo {volume} {30}},\ \bibinfo
  {pages} {100616} (\bibinfo {year} {2020})},\ \Eprint
  {http://arxiv.org/abs/2004.07867} {arXiv:2004.07867 [gr-qc]} \BibitemShut
  {NoStop}%
\bibitem [{\citenamefont {Ablowitz}\ \emph {et~al.}(1978)\citenamefont
  {Ablowitz}, \citenamefont {Ramani},\ and\ \citenamefont {Segur}}]{ars1}%
  \BibitemOpen
  \bibfield  {author} {\bibinfo {author} {\bibfnamefont {M.J.}\ \bibnamefont
  {Ablowitz}}, \bibinfo {author} {\bibfnamefont {A.}~\bibnamefont {Ramani}}, \
  and\ \bibinfo {author} {\bibfnamefont {H.}~\bibnamefont {Segur}},\ }\bibfield
   {title} {\enquote {\bibinfo {title} {{Nonlinear evolution equations and
  ordinary differential equations of painlev\'e type}},}\ }\href {\doibase
  https://doi.org/10.1007/BF02824479} {\bibfield  {journal} {\bibinfo
  {journal} {Lett. Nuovo Cimento}\ }\textbf {\bibinfo {volume} {23}},\ \bibinfo
  {pages} {333} (\bibinfo {year} {1978})}\BibitemShut {NoStop}%
\bibitem [{\citenamefont {Ablowitz}\ \emph
  {et~al.}(1980{\natexlab{a}})\citenamefont {Ablowitz}, \citenamefont
  {Ramani},\ and\ \citenamefont {Segur}}]{ars2}%
  \BibitemOpen
  \bibfield  {author} {\bibinfo {author} {\bibfnamefont {M.~J.}\ \bibnamefont
  {Ablowitz}}, \bibinfo {author} {\bibfnamefont {A.}~\bibnamefont {Ramani}}, \
  and\ \bibinfo {author} {\bibfnamefont {H.}~\bibnamefont {Segur}},\ }\bibfield
   {title} {\enquote {\bibinfo {title} {A connection between nonlinear
  evolution equations and ordinary differential equations of {P}-type. {I}},}\
  }\href {\doibase 10.1063/1.524491} {\bibfield  {journal} {\bibinfo  {journal}
  {Journal of Mathematical Physics}\ }\textbf {\bibinfo {volume} {21}},\
  \bibinfo {pages} {715--721} (\bibinfo {year}
  {1980}{\natexlab{a}})}\BibitemShut {NoStop}%
\bibitem [{\citenamefont {Ablowitz}\ \emph
  {et~al.}(1980{\natexlab{b}})\citenamefont {Ablowitz}, \citenamefont
  {Ramani},\ and\ \citenamefont {Segur}}]{ars3}%
  \BibitemOpen
  \bibfield  {author} {\bibinfo {author} {\bibfnamefont {M.~J.}\ \bibnamefont
  {Ablowitz}}, \bibinfo {author} {\bibfnamefont {A.}~\bibnamefont {Ramani}}, \
  and\ \bibinfo {author} {\bibfnamefont {H.}~\bibnamefont {Segur}},\ }\bibfield
   {title} {\enquote {\bibinfo {title} {A connection between nonlinear
  evolution equations and ordinary differential equations of {P}-type. {II}},}\
  }\href {\doibase 10.1063/1.524548} {\bibfield  {journal} {\bibinfo  {journal}
  {Journal of Mathematical Physics}\ }\textbf {\bibinfo {volume} {21}},\
  \bibinfo {pages} {1006--1015} (\bibinfo {year}
  {1980}{\natexlab{b}})}\BibitemShut {NoStop}%
\bibitem [{\citenamefont {Paliathanasis}\ and\ \citenamefont
  {Leach}(2016{\natexlab{b}})}]{Paliathanasis_2016}%
  \BibitemOpen
  \bibfield  {author} {\bibinfo {author} {\bibfnamefont {Andronikos}\
  \bibnamefont {Paliathanasis}}\ and\ \bibinfo {author} {\bibfnamefont
  {P.~G.~L.}\ \bibnamefont {Leach}},\ }\bibfield  {title} {\enquote {\bibinfo
  {title} {Nonlinear ordinary differential equations: A discussion on
  symmetries and singularities},}\ }\href {\doibase 10.1142/s0219887816300099}
  {\bibfield  {journal} {\bibinfo  {journal} {International Journal of
  Geometric Methods in Modern Physics}\ }\textbf {\bibinfo {volume} {13}},\
  \bibinfo {pages} {1630009} (\bibinfo {year}
  {2016}{\natexlab{b}})}\BibitemShut {NoStop}%
\bibitem [{\citenamefont {Lue}\ \emph {et~al.}(2004)\citenamefont {Lue},
  \citenamefont {Scoccimarro},\ and\ \citenamefont {Starkman}}]{Lue:2004rj}%
  \BibitemOpen
  \bibfield  {author} {\bibinfo {author} {\bibfnamefont {Arthur}\ \bibnamefont
  {Lue}}, \bibinfo {author} {\bibfnamefont {Roman}\ \bibnamefont
  {Scoccimarro}}, \ and\ \bibinfo {author} {\bibfnamefont {Glenn~D.}\
  \bibnamefont {Starkman}},\ }\bibfield  {title} {\enquote {\bibinfo {title}
  {{Probing Newton's constant on vast scales: DGP gravity, cosmic acceleration
  and large scale structure}},}\ }\href {\doibase 10.1103/PhysRevD.69.124015}
  {\bibfield  {journal} {\bibinfo  {journal} {Phys. Rev. D}\ }\textbf {\bibinfo
  {volume} {69}},\ \bibinfo {pages} {124015} (\bibinfo {year} {2004})},\
  \Eprint {http://arxiv.org/abs/astro-ph/0401515} {arXiv:astro-ph/0401515}
  \BibitemShut {NoStop}%
\bibitem [{\citenamefont {Linder}(2004)}]{Linder:2004ng}%
  \BibitemOpen
  \bibfield  {author} {\bibinfo {author} {\bibfnamefont {Eric~V.}\ \bibnamefont
  {Linder}},\ }\bibfield  {title} {\enquote {\bibinfo {title} {{Probing
  gravitation, dark energy, and acceleration}},}\ }\href {\doibase
  10.1103/PhysRevD.70.023511} {\bibfield  {journal} {\bibinfo  {journal} {Phys.
  Rev. D}\ }\textbf {\bibinfo {volume} {70}},\ \bibinfo {pages} {023511}
  (\bibinfo {year} {2004})},\ \Eprint {http://arxiv.org/abs/astro-ph/0402503}
  {arXiv:astro-ph/0402503} \BibitemShut {NoStop}%
\bibitem [{\citenamefont {Linder}\ and\ \citenamefont
  {Cahn}(2007)}]{Linder:2007hg}%
  \BibitemOpen
  \bibfield  {author} {\bibinfo {author} {\bibfnamefont {Eric~V.}\ \bibnamefont
  {Linder}}\ and\ \bibinfo {author} {\bibfnamefont {Robert~N.}\ \bibnamefont
  {Cahn}},\ }\bibfield  {title} {\enquote {\bibinfo {title} {{Parameterized
  Beyond-Einstein Growth}},}\ }\href {\doibase
  10.1016/j.astropartphys.2007.09.003} {\bibfield  {journal} {\bibinfo
  {journal} {Astropart. Phys.}\ }\textbf {\bibinfo {volume} {28}},\ \bibinfo
  {pages} {481--488} (\bibinfo {year} {2007})},\ \Eprint
  {http://arxiv.org/abs/astro-ph/0701317} {arXiv:astro-ph/0701317} \BibitemShut
  {NoStop}%
\bibitem [{\citenamefont {Basilakos}\ and\ \citenamefont
  {Pouri}(2012)}]{Basilakos:2012uu}%
  \BibitemOpen
  \bibfield  {author} {\bibinfo {author} {\bibfnamefont {Spyros}\ \bibnamefont
  {Basilakos}}\ and\ \bibinfo {author} {\bibfnamefont {Athina}\ \bibnamefont
  {Pouri}},\ }\bibfield  {title} {\enquote {\bibinfo {title} {{The growth index
  of matter perturbations and modified gravity}},}\ }\href {\doibase
  10.1111/j.1365-2966.2012.21168.x} {\bibfield  {journal} {\bibinfo  {journal}
  {Mon. Not. Roy. Astron. Soc.}\ }\textbf {\bibinfo {volume} {423}},\ \bibinfo
  {pages} {3761} (\bibinfo {year} {2012})},\ \Eprint
  {http://arxiv.org/abs/1203.6724} {arXiv:1203.6724 [astro-ph.CO]} \BibitemShut
  {NoStop}%
\bibitem [{\citenamefont {Steigerwald}\ \emph {et~al.}(2014)\citenamefont
  {Steigerwald}, \citenamefont {Bel},\ and\ \citenamefont
  {Marinoni}}]{Steigerwald:2014ava}%
  \BibitemOpen
  \bibfield  {author} {\bibinfo {author} {\bibfnamefont {Heinrich}\
  \bibnamefont {Steigerwald}}, \bibinfo {author} {\bibfnamefont {Julien}\
  \bibnamefont {Bel}}, \ and\ \bibinfo {author} {\bibfnamefont {Christian}\
  \bibnamefont {Marinoni}},\ }\bibfield  {title} {\enquote {\bibinfo {title}
  {{Probing non-standard gravity with the growth index: a background
  independent analysis}},}\ }\href {\doibase 10.1088/1475-7516/2014/05/042}
  {\bibfield  {journal} {\bibinfo  {journal} {JCAP}\ }\textbf {\bibinfo
  {volume} {05}},\ \bibinfo {pages} {042} (\bibinfo {year} {2014})},\ \Eprint
  {http://arxiv.org/abs/1403.0898} {arXiv:1403.0898 [astro-ph.CO]} \BibitemShut
  {NoStop}%
\bibitem [{\citenamefont {Papagiannopoulos}\ \emph {et~al.}(2017)\citenamefont
  {Papagiannopoulos}, \citenamefont {Basilakos}, \citenamefont {Paliathanasis},
  \citenamefont {Savvidou},\ and\ \citenamefont
  {Stavrinos}}]{Papagiannopoulos:2017whb}%
  \BibitemOpen
  \bibfield  {author} {\bibinfo {author} {\bibfnamefont {G.}~\bibnamefont
  {Papagiannopoulos}}, \bibinfo {author} {\bibfnamefont {S.}~\bibnamefont
  {Basilakos}}, \bibinfo {author} {\bibfnamefont {A.}~\bibnamefont
  {Paliathanasis}}, \bibinfo {author} {\bibfnamefont {S.}~\bibnamefont
  {Savvidou}}, \ and\ \bibinfo {author} {\bibfnamefont {P.~C.}\ \bibnamefont
  {Stavrinos}},\ }\bibfield  {title} {\enquote {\bibinfo {title}
  {{Finsler\textendash{}Randers cosmology: dynamical analysis and growth of
  matter perturbations}},}\ }\href {\doibase 10.1088/1361-6382/aa8be1}
  {\bibfield  {journal} {\bibinfo  {journal} {Class. Quant. Grav.}\ }\textbf
  {\bibinfo {volume} {34}},\ \bibinfo {pages} {225008} (\bibinfo {year}
  {2017})},\ \Eprint {http://arxiv.org/abs/1709.03748} {arXiv:1709.03748
  [gr-qc]} \BibitemShut {NoStop}%
\bibitem [{\citenamefont {Polarski}\ and\ \citenamefont
  {Gannouji}(2008)}]{Polarski:2007rr}%
  \BibitemOpen
  \bibfield  {author} {\bibinfo {author} {\bibfnamefont {David}\ \bibnamefont
  {Polarski}}\ and\ \bibinfo {author} {\bibfnamefont {Radouane}\ \bibnamefont
  {Gannouji}},\ }\bibfield  {title} {\enquote {\bibinfo {title} {{On the growth
  of linear perturbations}},}\ }\href {\doibase 10.1016/j.physletb.2008.01.032}
  {\bibfield  {journal} {\bibinfo  {journal} {Phys. Lett. B}\ }\textbf
  {\bibinfo {volume} {660}},\ \bibinfo {pages} {439--443} (\bibinfo {year}
  {2008})},\ \Eprint {http://arxiv.org/abs/0710.1510} {arXiv:0710.1510
  [astro-ph]} \BibitemShut {NoStop}%
\bibitem [{\citenamefont {Gannouji}\ and\ \citenamefont
  {Polarski}(2008)}]{Gannouji:2008jr}%
  \BibitemOpen
  \bibfield  {author} {\bibinfo {author} {\bibfnamefont {Radouane}\
  \bibnamefont {Gannouji}}\ and\ \bibinfo {author} {\bibfnamefont {David}\
  \bibnamefont {Polarski}},\ }\bibfield  {title} {\enquote {\bibinfo {title}
  {{The growth of matter perturbations in some scalar-tensor DE models}},}\
  }\href {\doibase 10.1088/1475-7516/2008/05/018} {\bibfield  {journal}
  {\bibinfo  {journal} {JCAP}\ }\textbf {\bibinfo {volume} {05}},\ \bibinfo
  {pages} {018} (\bibinfo {year} {2008})},\ \Eprint
  {http://arxiv.org/abs/0802.4196} {arXiv:0802.4196 [astro-ph]} \BibitemShut
  {NoStop}%
\bibitem [{\citenamefont {Frusciante}(2021)}]{Frusciante:2021sio}%
  \BibitemOpen
  \bibfield  {author} {\bibinfo {author} {\bibfnamefont {Noemi}\ \bibnamefont
  {Frusciante}},\ }\bibfield  {title} {\enquote {\bibinfo {title} {{Signatures
  of $f(Q)$-gravity in cosmology}},}\ }\href {\doibase
  10.1103/PhysRevD.103.044021} {\bibfield  {journal} {\bibinfo  {journal}
  {Phys. Rev. D}\ }\textbf {\bibinfo {volume} {103}},\ \bibinfo {pages}
  {044021} (\bibinfo {year} {2021})},\ \Eprint
  {http://arxiv.org/abs/2101.09242} {arXiv:2101.09242 [astro-ph.CO]}
  \BibitemShut {NoStop}%
\end{thebibliography}%

\end{document}